# DIFFUSION OF A FLUID THROUGH A VISCOELASTIC SOLID

SATISH KARRA

Abstract. This paper is concerned with the diffusion of a fluid through a viscoelastic solid undergoing large deformations. Using ideas from the classical theory of mixtures and a thermodynamic framework based on the notion of maximization of the rate of entropy production, the constitutive relations for a mixture of a viscoelastic solid and a fluid (specifically Newtonian fluid) are derived. By prescribing forms for the specific Helmholtz potential and the rate of dissipation, we derive the relations for the partial stress in the solid, the partial stress in the fluid, the interaction force between the solid and the fluid, and the evolution equation of the natural configuration of the solid. We also use the assumption that the volume of the mixture is equal to the sum of the volumes of the two constituents in their natural state as a constraint. Results from the developed model are shown to be in good agreement with the experimental data for the diffusion of various solvents through high temperature polyimides that are used in the aircraft industry. The swelling of a viscoelastic solid under the application of an external force is also studied.

## 1. Introduction

Several materials in the areas of of polymer mechanics, asphalt mechanics, and biomechanics that show non-linear viscoelastic behavior swell in the presence of a fluid. For instance, polyimides which are known to show viscoelastic solid-like response (see Bhargava (2007), Falcone and Ruggles-Wrenn (2009)) are used in the aerospace industry due to their good performance at high temperatures (also see Ghosh and Mittal (1996)); these materials in their service environment are known to swell in the presence of moisture. In addition, asphalt based materials (that are well know to show non-linear viscoelastic fluid-like behavior) degrade in the presence of moisture (Kim et al. (2004)). Diffusion of biological fluids through biological materials is another application wherein typically nutrition is provided by the fluid that diffuses, and the amount of the stress or strain in the solid can control the chemicals that are released (Rajagopal (2007)). Thus, there is a considerable interest to understand how such viscoelastic materials deform and swell due to diffusion of a fluid. Study of such a phenomenon is also of interest in geomechanics (Cohen (1992)) and food industry (Singh et al. (2004)).

It is well known that the Darcy's and Fick's equations (Darcy (1856); Fick (1855)) that are extensively used cannot predict swelling of the solid as well as the stresses in the solid. In fact, Darcy's equation is an approximation of the balance of linear momentum of the fluid going through a rigid solid. To capture the swelling phenomena, several works have been done using mixture theory (see review article by Rajagopal (2003)) and using variational principles (Baek and Srinivasa (2004)). These models have been shown to match well with experimental data for swelling of rubber (that show elastic response) due to the diffusion of various organic solvents.

In the area of diffusion of a fluid through viscoelastic materials, some of the earliest works were by Biot (1956) and Weitsman (1987), who have used linear viscoelasticity. In deriving their models, they have used the fact that the fluxes and affinities are related through linear phenomenonological relations. Later on, Cohen and co-workers (Cohen and White Jr. (1991); Edwards and Cohen (1995)), and Durning and co-workers (Huang and Durning (1997); Cairncross and Durning (1996)) have also recognized the importance of studying the diffusion of solvents through polymeric materials that show viscoelastic response. They have coupled diffusion and viscoelasticity by adding terms to the flux of the diffusing fluid that depend on the stress in the solid. Such an approach does not have a thermodynamic basis. Furthermore, these models developed are only one-dimensional in nature. Recently, Liu et al. (2005) have used the model developed by Cohen and co-workers to study the effect of various viscoelastic parameters on diffusion. They have shown that comparable relaxation times of polymer viscoelasticity and diffusion of a fluid results in non-Fickian behavior.

1.1. **Main contributions of this work.** Our main goal in this paper is to develop a thermodynamic framework to model diffusion of a fluid through a viscoelastic solid and we shall mainly focus on the swelling of polyimides. We use ideas from mixture theory (see Truesdell et al. (2004), Truesdell (1957a), Truesdell (1957b), Atkin and Craine (1976), Bedford and Drumheller (1983), Green and Naghdi (1969), Samohŷl (1987), Bowen (1982), Rajagopal and Tao (1995) for details) and irreversible thermodynamics to build such a framework. In Karra and Rajagopal (2010), a framework that can be used to predict the non-linear viscoelastic response of polyimides under various temperature and loading conditions, has been developed. In the current paper, such a framework in Karra and Rajagopal (2010) is extended to incorporate diffusion of a fluid and to model the swelling phenomenon.

The thermodynamic framework in the current work uses the notion of *evolving natural configuration* that has been used to model a variety of phenomena including classical plasticity, viscoelasticity, multi-network theory, superplasticity, twinning, etc. (see Rajagopal and Srinivasa (2004a) for details). The evolution of such a natural configuration is determined by maximizing the rate of entropy production (with any additional constraints). We constitutively prescribe forms for the Helmholtz potential of the mixture and the rate of dissipation (which is the product of density, temperature and the rate of entropy production) due to mechanical working, diffusion and heat conduction. The final constitutive relations are then derived by maximizing the rate of dissipation under appropriate constraints. In such an approach, one need not assume linear phenomenonological relations between the flux and the affinites, and thus our framework is more general. It has also been shown recently that if one chooses a quadratic form for the rate of entropy production in terms of affinities, and maximizes the rate of entropy production with respect to the affinities, one can arrive at the Onsager's relations (see Rajagopal and Srinivasa (2004b) for further details).

An initial boundary value problem is considered wherein a viscoelastic solid is held between two rigid walls, and immersed in a fluid. Using the model developed in the current paper, free swelling of such a solid and its swelling under the application of external force i.e., stress-assisted swelling are studied. The numerical results for free swelling of the viscoelastic solid are compared with experimental data for diffusion of different solvents through PMDA-ODA and HFPE-II-52 polyimide resins.

1.2. **Organization of the work.** In section (2), the kinematics required in this paper are documented. In section (3), the constitutive assumptions are specified and the constitutive relations are derived. We shall also show that our constitutive relations reduces to the equations for diffusion through an elastic solid derived using theory of mixtures when certain parameters take special values. An initial boundary value problem is set up using our model in section (4). The boundary conditions used, the non-dimensionalization scheme, and comparison of the numerical results with experimental data are given in (4.1), (4.2), and (4.3), respectively. Final concluding remarks are given in section (5).

## 2. Preliminaries

Let us consider a mixture of a Navier-Stokes fluid and a viscoelastic solid. We shall assume co-occupancy of the constituents, which is the central idea in theory of mixtures. It is based on the notion that at each point $\boldsymbol{x}$ in the mixture, at some time $t$, the two constituents exist together in a homogenized fashion, and are capable of moving relative to each other. We shall denote the quantities associated with the fluid through the superscript $f$ and use the superscript $s$ for the solid. Now, we shall define the motion $\boldsymbol{\chi}^i$ for the $i$-th constituent of the mixture through

$$\boldsymbol{x} = \boldsymbol{\chi}^i\left(\boldsymbol{X}^i, t\right), \quad i = f, s, \tag{2.1}$$

where $\boldsymbol{X}^i$ is the material point of the $i$-th constituent in its reference configuration. We shall assume that the mapping $\boldsymbol{\chi}^i$ is sufficiently smooth and invertible at each time $t$. The velocity associated with the $i$-th constituent is defined as

$$\boldsymbol{v}^i = \frac{\partial \boldsymbol{\chi}^i}{\partial t}, \tag{2.2}$$

and the deformation gradient through

$$\boldsymbol{F}^i = \frac{\partial \boldsymbol{\chi}^i}{\partial \boldsymbol{x}}. \tag{2.3}$$

Let $\kappa_t$ denote the current configuration of the mixture and let $\kappa_R$, $\kappa_r$ denote the reference configurations of the solid and the fluid respectively. Also, let $\kappa_{p(t)}$ denote the *natural configuration* of the viscoelastic solid (see Fig. (1)). Such a configuration is attained by the solid body upon instantaneous removal of external loading. For a Navier-Stokes fluid, the natural configuration is same as the current configuration of the fluid (Rajagopal (1995)). Now, if $\boldsymbol{F}^i$, $i = f, s$ is the gradient of the motion (usually known as deformation gradient) $\boldsymbol{\chi}^{\boldsymbol{i}}\left(\boldsymbol{X}^{\boldsymbol{i}}, t\right)$, and if $\boldsymbol{F}^s_{\kappa_{p(t)}}$ is the gradient of the motion of the viscoelastic solid from $\kappa_{p(t)}$ to $\kappa_t$, then

$$\boldsymbol{G}^s = \left(\boldsymbol{F}^s_{\kappa_{p(t)}}\right)^{-1} \boldsymbol{F}^s. \tag{2.4}$$

The density $\rho$ and the average velocity (also known as *barycentric* velocity) $\boldsymbol{v}$ of the mixture are defined by

$$\rho = \sum_i \rho^i, \quad \boldsymbol{v} = \frac{1}{\rho} \sum_i \rho^i \boldsymbol{v}^i. \tag{2.5}$$

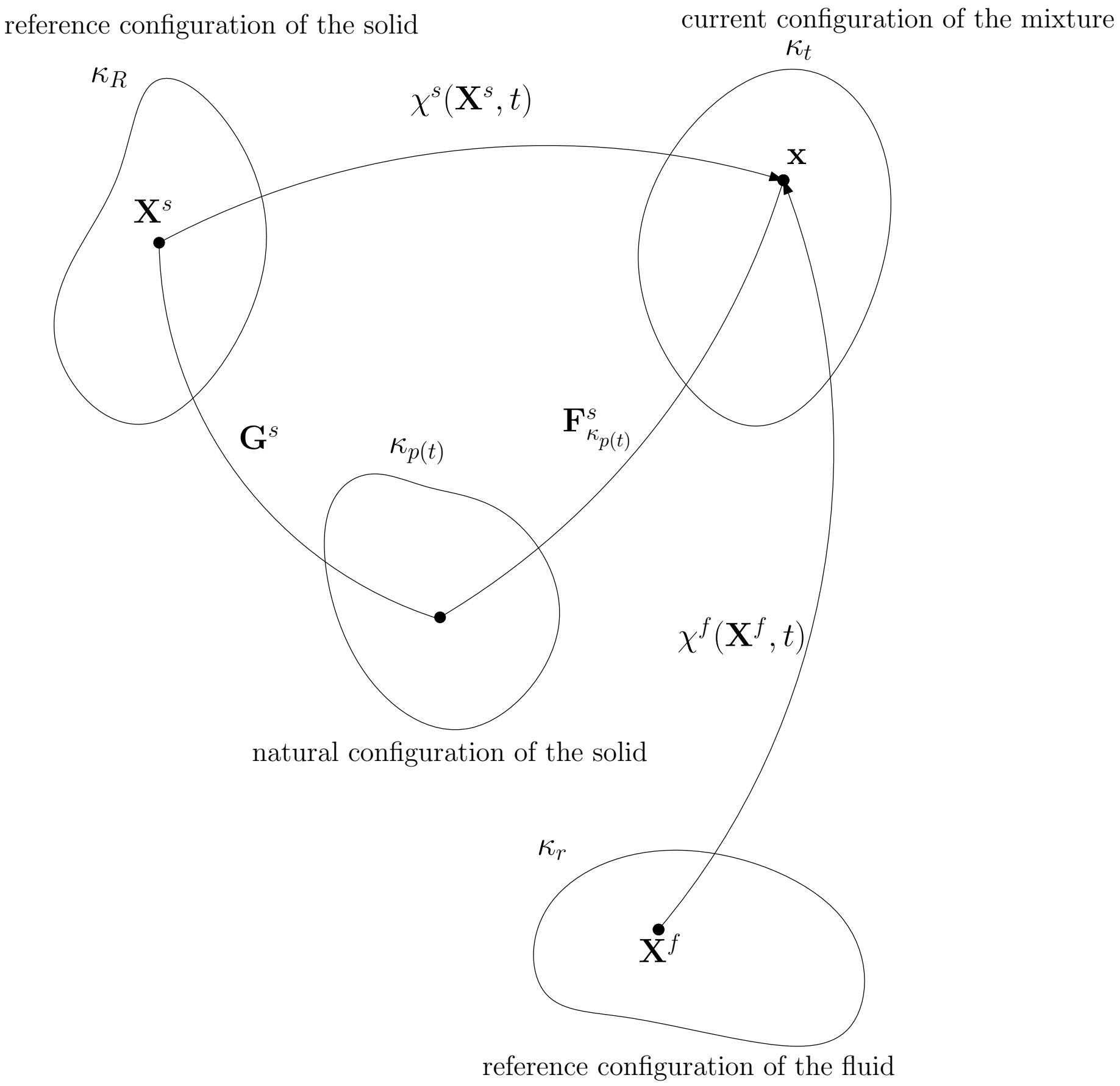

FIGURE 1. Illustration of the various configurations of the viscoelastic solid and fluid components in the mixture.

We define the following derivatives for any scalar quantity $\phi^i$ by

$$\frac{\partial \phi^i}{\partial t} = \frac{\partial \tilde{\phi}^i}{\partial t}, \quad \frac{d^i \phi^i}{dt} = \frac{\partial \hat{\phi}^i}{\partial t}, \quad \mathrm{grad}(\phi^i) = \frac{\partial \tilde{\phi}^i}{\partial \boldsymbol{x}}, \quad \frac{\partial \phi^i}{\partial \boldsymbol{X}^i} = \frac{\partial \hat{\phi}^i}{\partial \boldsymbol{X}^i}, \tag{2.6}$$

where

$$\phi^i = \tilde{\phi}^i\left(\boldsymbol{x}, t\right) = \hat{\phi}^i\left(\boldsymbol{X}^i, t\right). \tag{2.7}$$

Hence,

$$\frac{d^i \phi^i}{dt} = \frac{\partial \phi^i}{\partial t} + \mathrm{grad}(\phi^i) \cdot \boldsymbol{v}^i, \tag{2.8}$$

and we shall also define the following

$$\frac{d \phi}{dt} = \frac{\partial \phi}{\partial t} + \mathrm{grad}(\phi) \cdot \boldsymbol{v}. \tag{2.9}$$

The velocity gradient for the $i$-th component $\boldsymbol{L}^i$ and the velocity gradient for the total mixture $\boldsymbol{L}$ are defined by

$$\boldsymbol{L}^i = \mathrm{grad}(\boldsymbol{v}^i), \quad \boldsymbol{L} = \mathrm{grad}(\boldsymbol{v}). \tag{2.10}$$

The symmetric and anti-symmetric parts for the velocity gradients $\boldsymbol{L}^i$, $\boldsymbol{L}$ are

$$\boldsymbol{D}^i = \frac{1}{2}\left[\boldsymbol{L}^i + \left(\boldsymbol{L}^i\right)^T\right], \quad \boldsymbol{W}^i = \frac{1}{2}\left[\boldsymbol{L}^i - \left(\boldsymbol{L}^i\right)^T\right],$$
$$\boldsymbol{D} = \frac{1}{2}\left[\boldsymbol{L} + \left(\boldsymbol{L}\right)^T\right], \quad \boldsymbol{W} = \frac{1}{2}\left[\boldsymbol{L} - \left(\boldsymbol{L}\right)^T\right]. \tag{2.11}$$

The left Cauchy-Green stretch tensor $\boldsymbol{B}^s_p$, $\boldsymbol{B}^s_{p(t)}$ and their principal invariants are defined as

$$\boldsymbol{B}^s_G = \boldsymbol{F}^s_G \left(\boldsymbol{F}^s_G\right)^T, \quad \boldsymbol{B}^s_{p(t)} = \boldsymbol{F}^s_{\kappa_{p(t)}} \left(\boldsymbol{F}^s_{\kappa_{p(t)}}\right)^T, \tag{2.12}$$

$$\mathrm{I}_{B^s_j} = \mathrm{tr}\left(\boldsymbol{B}^s_j\right), \quad \mathrm{II}_{B^s_j} = \frac{1}{2}\left\{\left[\mathrm{tr}\left(\boldsymbol{B}^s_j\right)\right]^2 - \mathrm{tr}\left[\left(\boldsymbol{B}^s_j\right)^2\right]\right\}, \quad \mathrm{III}_{B^s_j} = \det\left(\boldsymbol{B}^s_j\right), \quad j = G, p(t), \tag{2.13}$$

where det(.) is the determinant of a second order tensor. We shall next note the balance laws.

The balance of mass for the $i$-th constituent without any mass production is given by

$$\frac{\partial \rho^i}{\partial t} + \mathrm{div}\left(\rho^i \boldsymbol{v}^i\right) = 0, \tag{2.14}$$

where $\rho^i$ is the mass density of the $i$-th constituent and div(.) := tr(grad(.)) is the divergence operator with tr(.) meaning the trace of a second order tensor. The summation of Eq. (2.14) over $i$ along with Eq. (2.5) leads to

$$\frac{\partial \rho}{\partial t} + \mathrm{div}\left(\rho \boldsymbol{v}\right) = 0. \tag{2.15}$$

The balance of linear momentum for $i$-th constituent is

$$\rho^i \frac{d^i \boldsymbol{v}^i}{dt} = \mathrm{div}\left[\left(\boldsymbol{T}^i\right)^T\right] + \rho^i \boldsymbol{b}^i + \boldsymbol{m}^i, \tag{2.16}$$

where $\boldsymbol{m}^i$ is the interaction force on the $i$-th constituent due to the other constituents, $\boldsymbol{b}^i$ is the external body force on the $i$-th constituent, $\boldsymbol{T}^i$ is the partial Cauchy stress tensor associated with the $i$-th constituent related to the surface traction on the $i$-th constituent $\boldsymbol{t}^i$ through

$$\boldsymbol{t}^i = \left(\boldsymbol{T}^i\right)^T \boldsymbol{n}, \tag{2.17}$$

where $\boldsymbol{n}$ is the surface outward normal. From Newton's third law, we have

$$\sum_i \boldsymbol{m}^i = 0. \tag{2.18}$$

For mixtures, the balance of angular momentum, in the absence of body couples requires that the total Cauchy stress of the mixture be symmetric i.e.,

$$\boldsymbol{T} = \boldsymbol{T}^T, \quad \text{where} \quad \boldsymbol{T} = \sum_i \boldsymbol{T}^i, \tag{2.19}$$

although the individual partial stresses $\boldsymbol{T}^i$ could be non-symmetric. Now, the balance of energy for the $i$-th constituent is given by

$$\rho^i \frac{d}{dt}\left(\epsilon^i + \frac{\boldsymbol{v}^i . \boldsymbol{v}^i}{2}\right) = \mathrm{div}\left(\boldsymbol{T}^i \cdot \boldsymbol{v}^i - \boldsymbol{q}^i\right) + \rho^i r^i + \rho^i \boldsymbol{b}^i \cdot \boldsymbol{v}^i + E^i + \boldsymbol{m}^i \cdot \boldsymbol{v}^i, \tag{2.20}$$

where $\epsilon^i$, $\boldsymbol{q}^i$, $r^i$ are the specific internal energy, heat flux, radiant heating associated with the $i$-th component and $E^i$ is the energy supplied to the $i$-th constituent from the other constituents.

Now, taking the scalar multiplication of Eq. (2.16) with $\boldsymbol{v}^i$ and subtracting the resulting equation from Eq. (2.20), we arrive at

$$\rho^i \frac{d\epsilon^i}{dt} = \boldsymbol{T}^i \cdot \boldsymbol{L}^i - \mathrm{div}(\boldsymbol{q}^i) + \rho^i r^i + E^i, \tag{2.21}$$

Using $\epsilon^i = \psi^i + \theta\eta^i$, where $\psi^i$, $\eta^i$ are the Helmholtz potential and specific entropy of the $i$-th constituent, and with $\theta$ being the common temperature of the constituents at a point in the mixture, Eq. (2.21) along with Eq. (2.14) results in

$$\begin{aligned} \frac{\partial}{\partial t}\left(\rho^i\eta^i\right) + \mathrm{div}\left(\rho^i\eta^i\boldsymbol{v}^i\right) = \frac{1}{\theta}\boldsymbol{T}^i \cdot \boldsymbol{L}^i - \mathrm{div}\left(\frac{\boldsymbol{q}^i}{\theta}\right) - \frac{1}{\theta^2}\boldsymbol{q}^i \cdot \mathrm{grad}(\theta) + \frac{1}{\theta}\rho^i r^i + \frac{1}{\theta}E^i \\ - \frac{\rho^i}{\theta}\left(\frac{d^i\psi^i}{dt} + \eta^i\frac{d^i\theta}{dt}\right). \end{aligned} \tag{2.22}$$

Now, using the fact that $\eta^i = -\frac{\partial\psi^i}{\partial\theta}$, we can establish the following result:

$$\begin{aligned} \frac{d^i\psi^i}{dt} + \eta^i\frac{d^i\theta}{dt} = \frac{d^i\psi^i}{dt} - \frac{\partial\psi^i}{\partial\theta}\frac{d^i\theta}{dt} = \left(\frac{\partial\psi^i}{\partial t} - \frac{\partial\psi^i}{\partial\theta}\frac{\partial\theta}{\partial t}\right) + \boldsymbol{v}^i \cdot \left(\mathrm{grad}(\psi^i) - \frac{\partial\psi^i}{\partial\theta}\mathrm{grad}(\theta)\right) \\ = \left(\frac{d^i\psi^i}{dt}\right)_{\theta\,\mathrm{fixed}}, \end{aligned} \tag{2.23}$$

where the subscript "$\theta$ fixed" means that the derivative is to be taken keeping $\theta$ fixed. We shall define

$$\boldsymbol{q} = \sum_i \boldsymbol{q}^i, \quad r = \frac{1}{\rho}\sum_i \rho^i r^i. \tag{2.24}$$

Using the relation Eq. (2.23) in Eq. (2.22) and summing over $i$, along with Eq. (2.24), we get

$$\begin{aligned} \frac{\partial}{\partial t}\left(\sum_i \rho^i\eta^i\right) + \mathrm{div}\left(\sum_i \rho^i\eta^i\boldsymbol{v}^i\right) = \frac{1}{\theta}\sum_i \boldsymbol{T}^i.\boldsymbol{L}^i - \mathrm{div}\left(\frac{\boldsymbol{q}}{\theta}\right) - \frac{1}{\theta^2}\boldsymbol{q} \cdot \mathrm{grad}(\theta) + \rho\left(\frac{r}{\theta}\right) \\ + \frac{1}{\theta}\sum_i E^i - \frac{1}{\theta}\sum_i \rho^i\left(\frac{d^i\psi^i}{dt}\right)_{\theta\,\mathrm{fixed}}. \end{aligned} \tag{2.25}$$

Eq. (2.25) is the balance of entropy with the rate of entropy production $\zeta$ being

$$\zeta = \frac{1}{\theta}\sum_i \boldsymbol{T}^i \cdot \boldsymbol{L}^i - \frac{1}{\theta^2}\boldsymbol{q} \cdot \mathrm{grad}(\theta) + \frac{1}{\theta}\sum_i E^i - \frac{1}{\theta}\sum_i \rho^i\left(\frac{d^i\psi^i}{dt}\right)_{\theta\,\mathrm{fixed}}. \tag{2.26}$$

We shall assume that the total entropy production can be additively split into entropy production due to thermal effects i.e., conduction ($\zeta_c$), and entropy production due to internal working and mixing ($\zeta_m$). We shall also require that each of these quantities be non-negative, so that the rate of entropy production $\zeta$ is non-negative and the second law of thermodynamics is satisfied automatically. This implies that

$$\zeta_c := -\frac{\boldsymbol{q}\cdot\mathrm{grad}(\theta)}{\theta^2} \geq 0, \tag{2.27a}$$

$$\zeta_m := \frac{1}{\theta}\sum_i \boldsymbol{T}^i\cdot\boldsymbol{L}^i + \frac{1}{\theta}\sum_i E^i - \frac{1}{\theta}\sum_i \rho^i\left(\frac{d^i\psi^i}{dt}\right)_{\theta\,\mathrm{fixed}} \geq 0. \tag{2.27b}$$

We shall choose $\boldsymbol{q} = -k(\rho,\theta)\mathrm{grad}(\theta),\ k \geq 0$, so that Eq. (2.27b) automatically satisfies. Also, if we define the rate of dissipation $\xi_m := \theta\zeta_m$, then

$$\sum_i \boldsymbol{T}^i\cdot\boldsymbol{L}^i + \sum_i E^i - \sum_i \rho^i\left(\frac{d^i\psi^i}{dt}\right)_{\theta\,\mathrm{fixed}} = \xi_m \geq 0. \tag{2.28}$$

Assuming

$$\sum_i E^i + \sum_i \boldsymbol{m}^i\cdot\boldsymbol{v}^i = 0, \tag{2.29}$$

Eq. (2.28) can be re-written as

$$\sum_i \boldsymbol{T}^i\cdot\boldsymbol{L}^i - \sum_i \boldsymbol{m}^i\cdot\boldsymbol{v}^i - \sum_i \rho^i\left(\frac{d^i\psi^i}{dt}\right)_{\theta\,\mathrm{fixed}} = \xi_m. \tag{2.30}$$

Now,

$$\begin{aligned}\sum_i \rho^i\left(\frac{d^i\psi^i}{dt}\right) &= \frac{\partial}{\partial t}\left(\sum_i \rho^i\psi^i\right) + \mathrm{div}\left(\sum_i \rho^i\psi^i\boldsymbol{v}^i\right)\\ &= \rho\frac{d\psi}{dt} + \mathrm{div}\left(\sum_i \rho^i\psi^i\left(\boldsymbol{v}^i-\boldsymbol{v}\right)\right),\end{aligned} \tag{2.31}$$

where $\psi := \frac{1}{\rho}\sum_i \rho^i\psi^i$ is the average Helmholtz potential of the mixture.

Finally, from Eqs. (2.31) and (2.30), we arrive at

$$\sum_i \boldsymbol{T}^i\cdot\boldsymbol{L}^i - \sum_i \boldsymbol{m}^i\cdot\boldsymbol{v}^i - \left[\rho\frac{d\psi}{dt} + \mathrm{div}\left(\sum_i \rho^i\psi^i\left(\boldsymbol{v}^i-\boldsymbol{v}\right)\right)\right]_{\theta\,\mathrm{fixed}} = \xi_m. \tag{2.32}$$

Assuming that all the components have the same Helmholtz potential Eq. (2.32) reduces to

$$\sum_i \boldsymbol{T}^i\cdot\boldsymbol{L}^i - \sum_i \boldsymbol{m}^i\cdot\boldsymbol{v}^i - \left(\rho\frac{d\psi}{dt}\right)_{\theta\,\mathrm{fixed}} = \xi_m, \tag{2.33}$$

where we have used Eq. (2.5). The second law of thermodynamics is invoked by ensuring $\xi_m \geq 0$. The preliminaries discussed so far are sufficient for the derivation of the constitutive equations in section (3).

## 3. Constitutive assumptions

We shall assume that the specific Helmholtz potential for the mixture is of the form

$$\psi = \hat{\psi}\left(\theta, \mathrm{I}_{B^s_G}, \mathrm{II}_{B^s_G}, \mathrm{III}_{B^s_G}, \mathrm{I}_{B^s_{p(t)}}, \mathrm{II}_{B^s_{p(t)}}, \mathrm{III}_{B^s_{p(t)}}\right), \tag{3.1}$$

and so

$$\frac{d\psi}{dt} = \frac{d^s\psi}{dt} + (\boldsymbol{v} - \boldsymbol{v}^s)\cdot \mathrm{grad}(\psi) \tag{3.2}$$

$$\begin{aligned}
&\Rightarrow \left(\frac{d\psi}{dt}\right)_{\theta\,\mathrm{fixed}} = \\
&\left[\left(\frac{\partial\hat{\psi}}{\partial \mathrm{I}_{B^s_{p(t)}}} + \mathrm{I}_{B^s_{p(t)}}\frac{\partial\hat{\psi}}{\partial \mathrm{II}_{B^s_{p(t)}}}\right)\boldsymbol{I} - \frac{\partial\hat{\psi}}{\partial \mathrm{II}_{B^s_{p(t)}}}\boldsymbol{B}^s_{p(t)} + \mathrm{III}_{B^s_{p(t)}}\frac{\partial\hat{\psi}}{\partial \mathrm{III}_{B^s_{p(t)}}}(\boldsymbol{B}^s_{p(t)})^{-1}\right]\cdot \dot{\boldsymbol{B}}^s_{p(t)} \\
&\quad + \left[\left(\frac{\partial\hat{\psi}}{\partial \mathrm{I}_{B^s_G}} + \mathrm{I}_{B^s_G}\frac{\partial\hat{\psi}}{\partial \mathrm{II}_{B^s_G}}\right)\boldsymbol{I} - \frac{\partial\hat{\psi}}{\partial \mathrm{II}_{B^s_G}}\boldsymbol{B}^s_G + \mathrm{III}_{B^s_G}\frac{\partial\hat{\psi}}{\partial \mathrm{III}_{B^s_G}}(\boldsymbol{B}^s_G)^{-1}\right]\cdot \dot{\boldsymbol{B}}^s_G \\
&\quad + (\boldsymbol{v} - \boldsymbol{v}^s)\cdot(\mathrm{grad}(\psi))_{\theta\,\mathrm{fixed}},
\end{aligned} \tag{3.3}$$

where $\dot{(\ )}$ is $\dfrac{d^s(\ )}{dt}$ for the sake of convenience.

Now,

$$\begin{aligned}
&\dot{\boldsymbol{F}}^s = \dot{\boldsymbol{F}}^s_{\kappa_{p(t)}}\boldsymbol{G}^s + \boldsymbol{F}^s_{\kappa_{p(t)}}\dot{\boldsymbol{G}}^s \\
&\Rightarrow \dot{\boldsymbol{F}}^s(\boldsymbol{F}^s)^{-1} = \dot{\boldsymbol{F}}^s_{\kappa_{p(t)}}\boldsymbol{G}^s(\boldsymbol{G}^s)^{-1}(\boldsymbol{F}^s_{\kappa_{p(t)}})^{-1} + \boldsymbol{F}^s_{\kappa_{p(t)}}\dot{\boldsymbol{G}}^s(\boldsymbol{G}^s)^{-1}(\boldsymbol{F}^s_{\kappa_{p(t)}})^{-1} \\
&\Rightarrow \boldsymbol{L}^s = \boldsymbol{L}^s_{p(t)} + \boldsymbol{F}^s_{\kappa_{p(t)}}\boldsymbol{L}^s_G(\boldsymbol{F}^s_{\kappa_{p(t)}})^{-1} \\
&\Rightarrow \boldsymbol{D}^s = \boldsymbol{D}^s_{p(t)} + \frac{1}{2}\left[\boldsymbol{F}^s_{\kappa_{p(t)}}\boldsymbol{L}^s_G(\boldsymbol{F}^s_{\kappa_{p(t)}})^{-1} + (\boldsymbol{F}^s_{\kappa_{p(t)}})^{-T}(\boldsymbol{L}^s_G)^T(\boldsymbol{F}^s_{\kappa_{p(t)}})^T\right].
\end{aligned} \tag{3.4}$$

In addition,

$$\begin{aligned}
\dot{\boldsymbol{B}}^s_{p(t)} &= \dot{\boldsymbol{F}}^s_{\kappa_{p(t)}}(\boldsymbol{F}^s)^T + \boldsymbol{F}^s(\dot{\boldsymbol{F}}^s_{\kappa_{p(t)}})^T \\
&= \boldsymbol{L}^s_{p(t)}\boldsymbol{B}^s_{p(t)} + \boldsymbol{B}^s_{p(t)}(\boldsymbol{L}^s_{p(t)})^T,
\end{aligned} \tag{3.5}$$

and similarly

$$\dot{\boldsymbol{B}}^s_G = \boldsymbol{L}^s_G\boldsymbol{B}^s_G + \boldsymbol{B}^s_G(\boldsymbol{L}^s_G)^T. \tag{3.6}$$

Assuming that the response of the viscoelastic solid from the current configuration to its natural configuration is isotropic elastic, we choose $\kappa_{p(t)}$ such that

$$\boldsymbol{F}^s_{\kappa_{p(t)}} = \boldsymbol{V}^s_{\kappa_{p(t)}}, \tag{3.7}$$

where $\boldsymbol{V}^s_{\kappa_{p(t)}}$ is the right stretch tensor in the polar decomposition of $\boldsymbol{F}^s_{\kappa_{p(t)}}$.

Using Eqs. (3.4), (3.5), (3.6), (3.7) in Eq. (3.3), we get

$$\left(\frac{d\psi}{dt}\right)_{\theta\,\text{fixed}} =$$

$$2\left[\left(\frac{\partial\hat{\psi}}{\partial \mathrm{I}^s_{B_{p(t)}}} + \mathrm{I}_{B^s_{p(t)}}\frac{\partial\hat{\psi}}{\partial \mathrm{II}_{B^s_{p(t)}}}\right)\boldsymbol{B}^s_{p(t)} - \frac{\partial\hat{\psi}}{\partial \mathrm{II}_{B^s_{p(t)}}}(\boldsymbol{B}^s_{p(t)})^2 + \mathrm{III}_{B^s_{p(t)}}\frac{\partial\hat{\psi}}{\partial \mathrm{III}^s_{B_{p(t)}}}\boldsymbol{I}\right]\cdot(\boldsymbol{L}^s - \boldsymbol{L}^s_G)$$

$$+2\left[\left(\frac{\partial\hat{\psi}}{\partial \mathrm{I}^s_{B_G}} + \mathrm{I}_{B^s_G}\frac{\partial\hat{\psi}}{\partial \mathrm{II}_{B^s_G}}\right)\boldsymbol{B}^s_G - \frac{\partial\hat{\psi}}{\partial \mathrm{II}^s_{B_G}}(\boldsymbol{B}^s_G)^2 + \mathrm{III}_{B^s_G}\frac{\partial\hat{\psi}}{\partial \mathrm{III}_{B^s_G}}\boldsymbol{I}\right]\cdot\boldsymbol{L}^s_G$$

$$+(\boldsymbol{v} - \boldsymbol{v}^s)\cdot(\mathrm{grad}(\psi))_{\theta\,\text{fixed}}\,, \tag{3.8}$$

In what follows, we shall assume that the reference configurations (subscript o) of the constituents are same as their natural states (subscript R) and so $\phi^i := \dfrac{\rho^i}{\rho^i_R} = \dfrac{\rho^i_o}{\det\boldsymbol{F}^i\rho^i_R} = \dfrac{1}{\det\boldsymbol{F}^i}$, $i = s, f$, where we have used the fact that $\rho^i_o = \rho^i_R$. This need not be true in general.

We shall also assume the volume additivity constraint that is based on the fact that the volume of the swollen viscoelastic solid is equal to the sum of the volumes of the unswollen viscoelastic solid and the fluid (Mills (1966)). In our case this constraint is given by,

$$\phi^s + \phi^f = 1, \tag{3.9}$$

and so Eq. (2.14) can be re-written as

$$\frac{\partial\phi^i}{\partial t} + \mathrm{div}\left(\phi^i\boldsymbol{v}^i\right) = 0, \tag{3.10}$$

which implies

$$\frac{\partial\sum_i\phi^i}{\partial t} + \mathrm{div}\left(\sum_i\phi^i\boldsymbol{v}^i\right) = 0 \tag{3.11}$$

$$\Rightarrow \mathrm{div}\left(\phi^f\boldsymbol{v}^f + \phi^s\boldsymbol{v}^s\right) = 0 \quad \text{(using Eq. (3.9))}. \tag{3.12}$$

Eq. (3.12) can be re-written as

$$\phi^s\mathrm{tr}(\boldsymbol{L}^s) + \phi^f\mathrm{tr}(\boldsymbol{L}^f) + \boldsymbol{v}^s\cdot\mathrm{grad}(\phi^s) + \boldsymbol{v}^f\cdot\mathrm{grad}(\phi^f) = 0. \tag{3.13}$$

Again from Eq. (3.9), we have

$$\mathrm{grad}(\phi^s) + \mathrm{grad}(\phi^f) = 0, \tag{3.14}$$

and hence, using Eq. (3.14) in Eq. (3.13), we arrive at

$$\phi^s\mathrm{tr}(\boldsymbol{L}^s) + \phi^f\mathrm{tr}(\boldsymbol{L}^f) + \boldsymbol{v}_{s,f}\cdot\mathrm{grad}(\phi^s) = 0, \tag{3.15}$$

where $\boldsymbol{v}_{s,f} = \boldsymbol{v}^s - \boldsymbol{v}^f$, is the velocity of the solid with respect to the fluid.

Next, we shall assume that the rate of entropy production is of the form

$$\xi_m = \xi_m\left(\boldsymbol{L}^s_G, \boldsymbol{F}^s_{p(t)}, \boldsymbol{L}^f, \theta, \boldsymbol{v}_{s,f}\right), \tag{3.16}$$

and so Eq. (2.33) along with Eq. (2.18) reduces to

$$\boldsymbol{T}^s\cdot\boldsymbol{L}^s + \boldsymbol{T}^f\cdot\boldsymbol{L}^f - \boldsymbol{m}^s\cdot\boldsymbol{v}_{s,f} - \left(\rho\frac{d\psi}{dt}\right)_{\theta\,\text{fixed}} = \xi_m\left(\boldsymbol{L}^s_G, \boldsymbol{F}^s_{p(t)}, \boldsymbol{L}^f, \theta, \boldsymbol{v}_{s,f}\right). \tag{3.17}$$

Using Eq. (3.8) in Eq. (3.17), we get

$$\boldsymbol{T}^s\cdot\boldsymbol{L}^s+\boldsymbol{T}^f\cdot\boldsymbol{L}^f-\boldsymbol{m}^s\cdot\boldsymbol{v}_{s,f}-\boldsymbol{T}^s_{p(t)}\cdot(\boldsymbol{L}^s-\boldsymbol{L}^s_G)-\boldsymbol{T}^s_G\cdot\boldsymbol{L}^s_G-\rho(\boldsymbol{v}-\boldsymbol{v}^s)\cdot(\mathrm{grad}(\psi))_{\theta\,\mathrm{fixed}}$$
$$=\xi_m\left(\boldsymbol{L}^s_G,\boldsymbol{F}^s_{p(t)},\boldsymbol{L}^f,\theta,\boldsymbol{v}_{s,f}\right),\qquad(3.18)$$

where

$$\boldsymbol{T}^s_{p(t)}:=2\rho\left[\left(\frac{\partial\hat\psi}{\partial\mathrm{I}_{\boldsymbol{B}^s_{p(t)}}}+\mathrm{I}_{\boldsymbol{B}^s_{p(t)}}\frac{\partial\hat\psi}{\partial\mathrm{II}_{\boldsymbol{B}^s_{p(t)}}}\right)\boldsymbol{B}^s_{p(t)}-\frac{\partial\hat\psi}{\partial\mathrm{II}_{\boldsymbol{B}^s_{p(t)}}}(\boldsymbol{B}^s_{p(t)})^2+\mathrm{III}_{\boldsymbol{B}^s_{p(t)}}\frac{\partial\hat\psi}{\partial\mathrm{III}^s_{\boldsymbol{B}_{p(t)}}}\boldsymbol{I}\right],\qquad(3.19)$$

$$\boldsymbol{T}^s_G:=2\rho\left[\left(\frac{\partial\hat\psi}{\partial\mathrm{I}_{\boldsymbol{B}^s_G}}+\mathrm{I}_{\boldsymbol{B}^s_G}\frac{\partial\hat\psi}{\partial\mathrm{II}_{\boldsymbol{B}^s_G}}\right)\boldsymbol{B}^s_G-\frac{\partial\hat\psi}{\partial\mathrm{II}_{\boldsymbol{B}^s_G}}(\boldsymbol{B}^s_G)^2+\mathrm{III}_{\boldsymbol{B}^s_G}\frac{\partial\hat\psi}{\partial\mathrm{III}_{\boldsymbol{B}^s_G}}\boldsymbol{I}\right].\qquad(3.20)$$

Eq. (3.18) with the constraint Eq. (3.15) can be written as

$$\boldsymbol{T}^s\cdot\boldsymbol{L}^s+\boldsymbol{T}^f\cdot\boldsymbol{L}^f-\boldsymbol{m}^s\cdot\boldsymbol{v}_{s,f}-\boldsymbol{T}^s_{p(t)}\cdot(\boldsymbol{L}^s-\boldsymbol{L}^s_G)-\boldsymbol{T}^s_G\cdot\boldsymbol{L}^s_G-\rho(\boldsymbol{v}-\boldsymbol{v}^s)\cdot(\mathrm{grad}\psi)_{\theta\,\mathrm{fixed}}$$
$$+\lambda\left(\phi^s\mathrm{tr}(\boldsymbol{L}^s)+\phi^f\mathrm{tr}(\boldsymbol{L}^f)+\boldsymbol{v}_{s,f}\cdot\mathrm{grad}(\phi^s)\right)=\xi_m\left(\boldsymbol{L}^s_G,\boldsymbol{F}^s_{p(t)},\boldsymbol{L}^f,\theta,\boldsymbol{v}_{s,f}\right),\qquad(3.21)$$

where $\lambda$ is a Lagrange multiplier.

We shall further assume that the rate of dissipation can be additively split into the rate of dissipation due to mechanical working of the viscoelastic solid, the rate of dissipation due to the Navier-Stokes fluid and the rate of dissipation due to diffusion of the fluid, with specific forms as follows:

$$\xi_m\left(\boldsymbol{L}^s_G,\boldsymbol{F}^s_{p(t)},\boldsymbol{L}^f,\theta,\boldsymbol{v}_{s,f}\right)=\xi\left(\boldsymbol{L}^s_G,\boldsymbol{B}^s_{p(t)},\theta\right)+\nu\boldsymbol{D}^f\cdot\boldsymbol{D}^f+\alpha(\theta)\boldsymbol{v}_{s,f}\cdot\boldsymbol{v}_{s,f}.\qquad(3.22)$$

Then, from Eq. (3.22) and Eq. (3.21), we arrive at

$$\boldsymbol{T}^s\cdot\boldsymbol{L}^s+\boldsymbol{T}^f\cdot\boldsymbol{L}^f-\boldsymbol{m}^s\cdot\boldsymbol{v}_{s,f}-\boldsymbol{T}^s_{p(t)}\cdot(\boldsymbol{L}^s-\boldsymbol{L}^s_G)-\boldsymbol{T}^s_G\cdot\boldsymbol{L}^s_G-\rho(\boldsymbol{v}-\boldsymbol{v}^s)\cdot(\mathrm{grad}(\psi))_{\theta\,\mathrm{fixed}}$$
$$+\lambda\left(\phi^s\mathrm{tr}(\boldsymbol{L}^s)+\phi^f\mathrm{tr}(\boldsymbol{L}^f)+\boldsymbol{v}_{s,f}\cdot\mathrm{grad}(\phi^s)\right)=\xi\left(\boldsymbol{L}^s_G,\boldsymbol{B}^s_{p(t)},\theta\right)+\nu\boldsymbol{D}^f\cdot\boldsymbol{D}^f+\alpha(\theta)\boldsymbol{v}_{s,f}\cdot\boldsymbol{v}_{s,f},\qquad(3.23)$$

which can be re-written as

$$\boldsymbol{L}^s\cdot\left[\boldsymbol{T}^s+\lambda\phi^s\boldsymbol{I}-\boldsymbol{T}^s_{p(t)}\right]+\boldsymbol{L}^f\cdot\left[\boldsymbol{T}^f+\lambda\phi^f\boldsymbol{I}-\nu\boldsymbol{D}^f\right]+\left(\boldsymbol{T}^s_{p(t)}-\boldsymbol{T}^s_G\right)\cdot\boldsymbol{L}^s_G$$
$$+\boldsymbol{v}_{s,f}\cdot\left[-\boldsymbol{m}^s+\lambda\,\mathrm{grad}(\phi^s)-\alpha(\theta)\boldsymbol{v}_{s,f}+\rho^f\,(\mathrm{grad}(\psi))_{\theta\,\mathrm{fixed}}\right]=\xi\left(\boldsymbol{L}^s_G,\boldsymbol{B}^s_{p(t)},\theta\right),\qquad(3.24)$$

using the fact that $\rho\left(\boldsymbol{v}-\boldsymbol{v}^s\right)=-\rho^f\boldsymbol{v}_{s,f}$. Since, the right hand side of Eq. (3.24) does not depend on $\boldsymbol{L}^s$, $\boldsymbol{L}^f$ and $\boldsymbol{v}_{s,f}$, we have

$$\boldsymbol{T}^s=-\lambda\phi^s\boldsymbol{I}+\boldsymbol{T}^s_{p(t)},\qquad(3.25)$$

$$\boldsymbol{T}^f=-\lambda\phi^f\boldsymbol{I}+\nu\boldsymbol{D}^f,\qquad(3.26)$$

$$\boldsymbol{m}^s=\lambda\,\mathrm{grad}(\phi^s)-\alpha(\theta)\boldsymbol{v}_{s,f}+\rho^f\,(\mathrm{grad}(\psi))_{\theta\,\mathrm{fixed}},\qquad(3.27)$$

and so Eq. (3.24) reduces to

$$\left(\boldsymbol{T}^s_{p(t)}-\boldsymbol{T}^s_G\right)\cdot\boldsymbol{L}^s_G=\xi\left(\boldsymbol{L}^s_G,\boldsymbol{B}^s_{p(t)},\theta\right).\qquad(3.28)$$

Next, we shall maximize the rate of dissipation $\xi$ with Eq. (3.28). We shall maximize the auxiliary function

$$\Phi:=\xi+\beta\left[\xi-\left(\boldsymbol{T}^s_{p(t)}-\boldsymbol{T}^s_G\right)\cdot\boldsymbol{L}^s_G\right].\qquad(3.29)$$

Now,

$$\frac{\partial \Phi}{\partial \boldsymbol{L}_G^s} = 0 \Rightarrow \frac{(\beta+1)}{\beta}\frac{\partial \xi}{\partial \boldsymbol{L}_G^s} - \left(\boldsymbol{T}_{p(t)}^s - \boldsymbol{T}_G^s\right) = 0. \tag{3.30}$$

Taking the scalar product of Eq. (3.30) with $\boldsymbol{L}_G^s$ and using Eq. (3.28), we arrive at

$$\frac{(\beta+1)}{\beta} = \frac{\xi}{\dfrac{\partial \xi}{\partial \boldsymbol{L}_G^s}\cdot \boldsymbol{L}_G^s}, \tag{3.31}$$

and hence the evolution equation for the natural configuration of the solid is given by

$$\left(\boldsymbol{T}_{p(t)}^s - \boldsymbol{T}_G^s\right) = \frac{\xi}{\dfrac{\partial \xi}{\partial \boldsymbol{L}_G^s}\cdot \boldsymbol{L}_G^s}\frac{\partial \xi}{\partial \boldsymbol{L}_G^s}. \tag{3.32}$$

3.1. **Specific constitutive assumptions.** We shall assume the following specific form for the specific Helmholtz potential of the mixture

$$\begin{aligned}\hat{\psi} = A^s + (B^s + c_2^s)(\theta - \theta_s) - \frac{c_1^s}{2}(\theta-\theta_s)^2 - c_2^s\theta \ln\left(\frac{\theta}{\theta_s}\right) + \frac{\mu_{G0}-\mu_{G1}\theta}{\rho^s\theta_s}(\mathrm{I}_{B_G^s}-3)\\ + \frac{\mu_{p0}-\mu_{p1}\theta}{\rho^s\theta_s}(\mathrm{I}_{B_{p(t)}^s}-3) + \frac{R\theta}{\rho_R^f V_0\phi^s}\left[(1-\phi^s)\ln(1-\phi^s) - \chi(\phi^s)^2\right],\end{aligned} \tag{3.33}$$

where $\mu_{G0}, \mu_{G1}, \mu_{p0}, \mu_{p1}$ are material parameters, $\theta_s$ is a reference temperature for the viscoelastic solid, $R$ is the gas constant, $\theta$ the absolute temperature of mixture, $V_0$ the molar volume of the fluid, and $\chi$ a mixing parameter for the particular solid-fluid combination. The last term in Eq. (3.33) is the term added to the specific Helmholtz in (Karra and Rajagopal (2010)) to capture the swelling phenomenon in the solid.

Now,

$$\begin{aligned}\eta &= -\frac{\partial \hat{\psi}}{\partial \theta}\\ &= -(B^s + c_2^s) + c_1^s(\theta-\theta_s) + c_2^s \ln\left(\frac{\theta}{\theta_s}\right) + c_2^s + \frac{\mu_{G1}}{\rho\theta_s}(\mathrm{I}_{B_G^s}-3) + \frac{\mu_{p1}}{\rho\theta_s}(\mathrm{I}_{B_{p(t)}^s}-3)\\ &\quad - \frac{R}{\rho_R^f V_0\phi^s}\left[(1-\phi^s)\ln(1-\phi^s) - \chi(\phi^s)^2\right].\end{aligned} \tag{3.34}$$

The internal energy $\epsilon$ is given by

$$\begin{aligned}\epsilon &= \psi + \theta\eta\\ &= A^s - B^s\theta_s + c_2^s(\theta-\theta_s) + \frac{c_1^s}{2}\left(\theta^2-\theta_s^2\right) + \frac{\mu_{G0}}{\rho\theta_s}(\mathrm{I}_{B_G^s}-3) + \frac{\mu_{p0}}{\rho\theta_s}(\mathrm{I}_{B_{p(t)}^s}-3),\end{aligned} \tag{3.35}$$

and the specific heat capacity $C_v$ is

$$C_v = \frac{\partial \epsilon}{\partial \theta} = c_1^s\theta + c_2^s. \tag{3.36}$$

From Eq. (3.33) and Eq. (3.19)

$$\boldsymbol{T}^s_{p(t)} = \frac{\rho J^s_p J^s_G}{\rho^s_R}\left[2\bar{\mu}_p \boldsymbol{B}^s_{p(t)} + \bar{\mu}_p \left(\mathrm{I}_{B^s_{p(t)}} - 3\right)\boldsymbol{I} + \bar{\mu}_G \left(\mathrm{I}_{B^s_G} - 3\right)\boldsymbol{I}\right] + \frac{\rho R\theta J^s_p J^s_G}{\rho^f_R V_0}\left[\ln(1-\phi^s) + \phi^s + \chi(\phi^s)^2\right]\boldsymbol{I}, \tag{3.37}$$

and from Eq. (3.33) and Eq. (3.20),

$$\boldsymbol{T}^s_G = \frac{\rho J^s_p J^s_G}{\rho^s_R}\left[2\bar{\mu}_G \boldsymbol{B}^s_G + \bar{\mu}_p \left(\mathrm{I}_{B^s_{p(t)}} - 3\right)\boldsymbol{I} + \bar{\mu}_G \left(\mathrm{I}_{B^s_G} - 3\right)\boldsymbol{I}\right] + \frac{\rho R\theta J^s_p J^s_G}{\rho^f_R V_0}\left[\ln(1-\phi^s) + \phi^s + \chi(\phi^s)^2\right]\boldsymbol{I}, \tag{3.38}$$

where $\bar{\mu}_G = \frac{\bar{\mu}_{G0} - \bar{\mu}_{G1}\theta}{\theta_s}$, $\bar{\mu}_p = \frac{\bar{\mu}_{p0} - \bar{\mu}_{p1}\theta}{\theta_s}$, $J^s_G = \det(\boldsymbol{G}^s)$, $J^s_p = \det(\boldsymbol{F}^s_{\kappa_{p(t)}})$.

We shall further assume that the rate of dissipation $\xi$ is of the form

$$\xi = \gamma(\theta)\boldsymbol{D}^s_G \cdot \boldsymbol{D}^s_G, \tag{3.39}$$

then Eq. (3.25) becomes

$$\boldsymbol{T}^s = -\lambda\phi^s\boldsymbol{I} + \frac{\rho}{\rho^s}\left[2\bar{\mu}_p \boldsymbol{B}_{p(t)} + \bar{\mu}_p \left(\mathrm{I}_{B^s_{p(t)}} - 3\right)\boldsymbol{I} + \bar{\mu}_G \left(\mathrm{I}_{B^s_G} - 3\right)\boldsymbol{I}\right] + \frac{\rho R\theta J^s_p J^s_G}{\rho^f_R V_0}\left[\ln(1-\phi^s) + \phi^s + \chi(\phi^s)^2\right]\boldsymbol{I}, \tag{3.40}$$

and Eq. (3.32) reduces to

$$\frac{2\rho}{\rho^s}\left[\bar{\mu}_p \boldsymbol{B}^s_{p(t)} - \bar{\mu}_G \boldsymbol{B}^s_G\right] = \gamma(\theta)\boldsymbol{D}^s_G. \tag{3.41}$$

The final constitutive equations are

$$\boldsymbol{T}^s = -\lambda\phi^s\boldsymbol{I} + \frac{\rho}{\rho^s}\left[2\bar{\mu}_p \boldsymbol{B}^s_{p(t)} + \bar{\mu}_p \left(\mathrm{I}_{B^s_{p(t)}} - 3\right)\boldsymbol{I} + \bar{\mu}_G \left(\mathrm{I}_{B^s_G} - 3\right)\boldsymbol{I}\right] + \frac{\rho R\theta J^s_p J^s_G}{\rho^f_R V_0}\left[\ln(1-\phi^s) + \phi^s + \chi(\phi^s)^2\right]\boldsymbol{I}, \tag{3.42a}$$

$$\boldsymbol{T}^f = -\lambda\phi^f\boldsymbol{I} + \nu\boldsymbol{D}^f, \tag{3.42b}$$

$$\boldsymbol{m}^s = \lambda\ \mathrm{grad}\phi^s - \alpha(\theta)\boldsymbol{v}_{s,f} + \rho^f\left(\mathrm{grad}\psi\right)_{\theta\,\mathrm{fixed}}, \tag{3.42c}$$

and

$$\frac{2\rho}{\rho^s}\left[\bar{\mu}_p \boldsymbol{B}^s_{p(t)} - \bar{\mu}_G \boldsymbol{B}^s_G\right] = \gamma(\theta)\boldsymbol{D}^s_G, \tag{3.43}$$

being the evolution equation of the natural configuration of the solid.

Notice that when $\bar{\mu}_G = 0$ and $\gamma \to \infty$, for the LHS of Eq. (3.43) to be finite, we must have $\boldsymbol{D}^s_G = 0$. This implies that $\boldsymbol{G}^s = \boldsymbol{I}$, and hence $\boldsymbol{B}^s_{p(t)} = \boldsymbol{B}^s$ and the solid is now an elastic solid. In such a case, the constitutive equations Eq. (3.42), with additional assumption of $\nu = 0$, reduce to

$$\boldsymbol{T}^s = -\lambda \phi^s \boldsymbol{I} + 2\rho \left[ \left( \frac{\partial \hat{\psi}}{\partial \mathrm{I}_{B^s}} + \mathrm{I}_{B^s} \frac{\partial \hat{\psi}}{\partial \mathrm{II}_{B^s}} \right) \boldsymbol{B}^s - \frac{\partial \hat{\psi}}{\partial \mathrm{II}_{B^s}} (\boldsymbol{B}^s)^2 + \mathrm{III}_{B^s} \frac{\partial \hat{\psi}}{\partial \mathrm{III}_{B^s}} \boldsymbol{I} \right], \tag{3.44a}$$

$$\boldsymbol{T}^f = -\lambda \phi^f \boldsymbol{I}, \tag{3.44b}$$

$$\boldsymbol{m}^s = \lambda \ \mathrm{grad} \phi^s - \alpha(\theta) \boldsymbol{v}_{s,f} + \rho^f \left( \mathrm{grad} \psi \right)_{\theta \, \mathrm{fixed}} . \tag{3.44c}$$

These equations are same as the equations derived using theory of mixtures for the diffusion of a fluid through an elastic solid (see equations 3.15–3.17 in (Hron et al. (2002))).

## 4. Initial boundary value problem

Let us consider the problem of compression of the viscoelastic solid body inside rigid walls as shown in Fig. (2). Let us assume that the motion of the swollen solid body is given by

$$x = X, \quad y = Y, \quad z = f(Z, t). \tag{4.1}$$

In this case, the deformation gradient of the solid ($\boldsymbol{F}^s$) is given by

$$\boldsymbol{F}^s = \mathrm{diag} \left\{ 1, 1, p \right\}, \tag{4.2}$$

where $p := \dfrac{\partial f}{\partial Z}$. Let us assume a form for $\boldsymbol{G}^s$ as follows

$$\boldsymbol{G}^s = \mathrm{diag} \left\{ 1, 1, g(Z, t) \right\}, \tag{4.3}$$

and the velocity of the fluid be of the form

$$\boldsymbol{v}^f = (0, 0, v(Z, t)) . \tag{4.4}$$

Then,

$$\boldsymbol{B}^s_{\kappa_{p(t)}} = \mathrm{diag} \left\{ 1, 1, \left( \frac{p}{g} \right)^2 \right\}, \tag{4.5}$$

$$\mathrm{I}_{B^s_{p(t)}} = 2 + \left( \frac{p}{g} \right)^2, \quad \mathrm{I}_{B^s_G} = 2 + g^2, \quad J^s_p = \frac{p}{g}, \quad J^s_G = g. \tag{4.6}$$

The balance of mass for the solid gives

$$\phi^s = \frac{1}{\det(\boldsymbol{F}^s)} = \frac{1}{p}, \tag{4.7}$$

and hence from volume additivity constraint

$$\rho^f = \rho^f_R \left( 1 - \frac{1}{p} \right). \tag{4.8}$$

Also, we note the following relations

$$\frac{\rho}{\rho^s} = 1 + \frac{\rho_R^f}{\rho_R^s}(p-1), \tag{4.9a}$$

$$\frac{\rho}{\rho_R^f} = 1 + \frac{1}{p}\left(\frac{\rho_R^s}{\rho_R^f} - 1\right). \tag{4.9b}$$

The balance of mass for the fluid reduces to

$$\frac{\partial p}{\partial t} + \frac{v}{p}\frac{\partial p}{\partial Z} + (p-1)\frac{\partial v}{\partial Z} = 0. \tag{4.10}$$

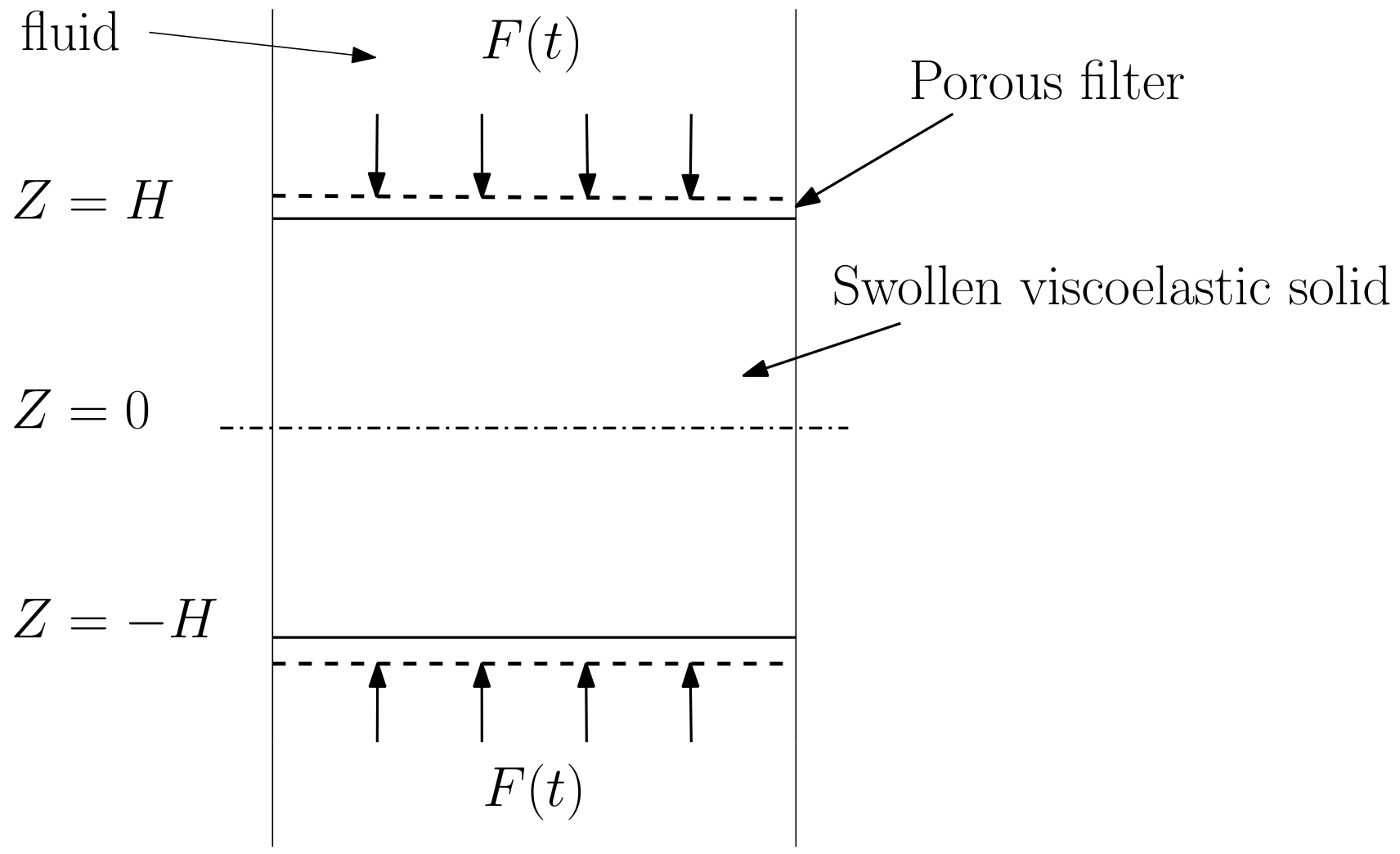

FIGURE 2. Schematic of the initial boundary value problem

Setting $\nu = 0$, $zz$-component of the total stress tensor for the mixture reduces to

$$T_{zz} = T_{zz}^s + T_{zz}^f \tag{4.11}$$

$$= -\lambda + \left(1 + \frac{\rho_R^f}{\rho_R^s}(p-1)\right)\left[2\bar{\mu}_p\left(\frac{p}{g}\right)^2 + \bar{\mu}_p\left(\frac{p^2}{g^2} - 1\right) + \bar{\mu}_G\left(g^2 - 1\right)\right]$$

$$+ \left(1 + \frac{1}{p}\left(\frac{\rho_R^s}{\rho_R^f} - 1\right)\right)\frac{R\theta p}{V_0}\left[\ln\left(1 - \frac{1}{p}\right) + \frac{1}{p} + \chi\frac{1}{p^2}\right]. \tag{4.12}$$

The balance of linear momentum for the solid and the fluid after assuming zero body forces reduce to

$$\rho^s\frac{\partial^2 f}{\partial t^2} = \frac{\partial T_{zz}^s}{\partial z} + m_z^s, \tag{4.13a}$$

$$\rho^f\frac{\partial v}{\partial t} = \frac{\partial T_{zz}^f}{\partial z} - m_z^s. \tag{4.13b}$$

Now, adding Eqs. (4.13a), (4.13b), we arrive at the balance of linear momentum for the mixture

$$\rho^s \frac{\partial^2 f}{\partial t^s} + \rho^f \frac{\partial v}{\partial t} = \frac{\partial}{\partial z}(T^s_{zz} + T^f_{zz})$$
$$\Rightarrow \rho^s \frac{\partial^2 f}{\partial t^s} + \rho^f \frac{\partial v}{\partial t} = -\frac{\partial \lambda}{\partial z} + \frac{\partial T^{sf}_{zz}}{\partial z}, \tag{4.14}$$

where

$$T^{sf}_{zz} = \left(1 + \frac{\rho^f_R}{\rho^s_R}(p-1)\right)\left[2\bar{\mu}_p\left(\frac{p}{g}\right)^2 + \bar{\mu}_p\left(\frac{p^2}{g^2} - 1\right) + \bar{\mu}_G\left(g^2 - 1\right)\right]$$
$$+ \left(1 + \frac{1}{p}\left(\frac{\rho^s_R}{\rho^f_R} - 1\right)\right)\frac{R\theta p}{V_0}\left[\ln\left(1 - \frac{1}{p}\right) + \frac{1}{p} + \chi\frac{1}{p^2}\right]. \tag{4.15}$$

Now, Eq. (4.13b) along with volume additivity constraint reduces to

$$\rho^f \frac{\partial v}{\partial t} = -\phi^f \frac{\partial \lambda}{\partial z} + \alpha\left(\frac{\partial f}{\partial t} - v\right) - \rho^f\left(\frac{\partial \psi}{\partial z}\right)_{\theta\,\text{fixed}}, \tag{4.16}$$

where we have used the fact that the velocity of the solid is $(0, 0, \frac{\partial f}{\partial t})$.

Multiplying Eq. (4.14) with $\phi^f$ and subtracting Eq. (4.16) from the resulting equation we get

$$\phi^f \rho^s \frac{\partial^2 f}{\partial t^2} + (\phi^f - 1)\rho^f \frac{\partial v}{\partial t} = -\alpha\left(\frac{\partial f}{\partial t} - v\right) + \phi^f \frac{\partial T^{sf}_{zz}}{\partial z} + \rho^f\left(\frac{\partial \psi}{\partial z}\right)_{\theta\,\text{fixed}}, \tag{4.17}$$

which reduces to

$$\phi^f \rho^s \frac{\partial^2 f}{\partial t^2} + (\phi^f - 1)\rho^f \frac{\partial v}{\partial t} = -\alpha\left(\frac{\partial f}{\partial t} - v\right) + \phi^f \frac{\partial T^{sf}_{zz}}{\partial z} + \rho^f\left(\frac{\partial \tilde{\psi}}{\partial z}\right), \tag{4.18}$$

where

$$\tilde{\psi} = \frac{\bar{\mu}_G p}{\rho^s_R}\left(g^2 - 1\right) + \frac{p\bar{\mu}_p}{\rho^s_R}\left(\frac{p^2}{g^2} - 1\right) + \frac{R\theta p}{\rho^f_R V_0}\left[\left(1 - \frac{1}{p}\right)\ln\left(1 - \frac{1}{p}\right) - \chi\frac{1}{p^2}\right]. \tag{4.19}$$

Now, assuming that the velocity and acceleration of the solid are small compared to that of the fluid, we shall drop $\frac{\partial f}{\partial t}$ and $\frac{\partial^2 f}{\partial t^2}$ in Eq. (4.18), we get

$$(\phi^f - 1)\rho^f \frac{\partial v}{\partial t} = \alpha v + \phi^f \frac{\partial T^{sf}_{zz}}{\partial z} + \rho^f\left(\frac{\partial \tilde{\psi}}{\partial z}\right). \tag{4.20}$$

Next, we shall also assume that the acceleration of the fluid is also small and we shall drop $\frac{\partial v}{\partial t}$ term in Eq. (4.20), to get

$$v = -\frac{1}{\alpha}\left[\phi^f \frac{\partial T^{sf}_{zz}}{\partial z} + \rho^f\left(\frac{\partial \tilde{\psi}}{\partial z}\right)\right]. \tag{4.21}$$

Using Eq. (4.21) in Eq. (4.10), we arrive at

$$\frac{\partial p}{\partial t} = \frac{1}{\alpha p^2}\frac{\partial p}{\partial Z}\left[\phi^f \frac{\partial T^{sf}_{zz}}{\partial Z} + \rho^f \left(\frac{\partial \tilde{\psi}}{\partial Z}\right)\right] + \frac{p-1}{\alpha}\frac{\partial}{\partial Z}\left[\frac{\phi^f}{p}\frac{\partial T^{sf}_{zz}}{\partial Z} + \frac{\rho^f}{p}\left(\frac{\partial \tilde{\psi}}{\partial Z}\right)\right]. \tag{4.22}$$

Also, note that

$$\boldsymbol{D}_G = \boldsymbol{L}_G = \operatorname{diag}\left\{0, 0, \frac{1}{g}\frac{\partial g}{\partial t}\right\}, \tag{4.23}$$

and so the evolution equation of the natural configuration reduces to

$$\gamma \frac{1}{g}\frac{\partial g}{\partial t} = 2\left(1 + \frac{\rho^f_R}{\rho^s_R}(p-1)\right)\left[\bar{\mu}_p \left(\frac{p}{g}\right)^2 - \bar{\mu}_G g^2\right]. \tag{4.24}$$

4.1. **Boundary conditions.** Applying boundary conditions in an initial boundary value problem has been an issue in mixture theory due to its basic assumption of co-occupancy. For instance, if traction is applied on the boundary, a natural question is how is the traction to be split between the solid and the fluid. To this end, the method of spitting the traction based on the volume fraction of the solid and the fluid was proposed (Rajagopal and Tao (1995)). Later on, Baek and Srinivasa (2004) derived the relations for swelling of an elastic body based on variational principles and the boundary conditions were derived *naturally.* However, this approach assumes that the swelling is *slow*, and that the relative velocity between the solid and the diffusing fluid is small. Recently, Prasad and Rajagopal (2006) have compared the solutions of diffusion of a fluid through a elastic slab using various boundary conditions like saturation boundary condition, traction splitting boundary condition, the natural boundary condition derived by Baek and Srinivasa, and the condition that the chemical potential is continuous across the boundary. Interestingly, they show that the results are *insensitive* to these different forms of boundary conditions.

For our problem, let $F(t)$ be the compressive force applied on the solid at $Z = \pm H$ as shown in Fig. (2) and let $P_\infty$ be the pressure in the fluid at the boundaries $Z = \pm H$, then we shall apply the following boundary conditions:

$$T^s_{zz} = -F(t) - \phi^s P_\infty, \quad Z = \pm H, \tag{4.25a}$$

$$T^f_{zz} = -\phi^f P_\infty, \quad Z = \pm H, \tag{4.25b}$$

that is, we are assuming that the external force is borne by the solid only, while the fluid pressure is borne by both the solid and the fluid, and this pressure is split proportional to the volume fraction of the constituents. Based on these assumptions, Eq. (4.25b) reduces to

$$\lambda = P_\infty, \quad Z = \pm H. \tag{4.26}$$

Eq. (4.25a) and Eq. (4.26) reduce to

$$-F(t) = T^{sf}_{zz}, \quad Z = \pm H. \tag{4.27}$$

Note that $F(t)$ is zero under free-swelling.

4.2. **Non-dimensionalization.** We shall use the following non-dimensionalization scheme:

$$Z^\star = \frac{Z}{L},\ t^\star = \frac{t}{T},\ v^\star = \frac{vT}{L},\ p^\star = p,\ g^\star = g,\ \bar{\mu}_p^\star = \frac{\bar{\mu}_p}{\mu},\ \gamma^\star = \frac{\gamma V}{\mu L}, \tag{4.28}$$

where $T$, $L$ are characteristic time and length respectively. If we pick $\mu = \frac{R\theta}{V_0}$ and define the non-dimensionalization quantities $\beta_1 := \frac{\rho_R^s}{\rho_R^f}$, $\beta_2 := \frac{L^2 V_0 \alpha}{R\theta T}$ then Eqs. (4.22), (4.24) become

$$\begin{aligned}\beta_2 \frac{\partial p\star}{\partial t} &= \frac{1}{(p^\star)^2}\left(1 - \frac{1}{p^\star}\right)\frac{\partial p^\star}{\partial Z^\star}\left[\frac{\partial T_{zz}^{sf\star}}{\partial Z^\star} + \frac{\partial \tilde{\psi}^\star}{\partial Z^\star}\right] \\ &\quad + (p^\star - 1)\frac{\partial}{\partial Z^\star}\left[\left(1 - \frac{1}{p^\star}\right)\frac{1}{p^\star}\left(\frac{\partial T_{zz}^{sf\star}}{\partial Z^\star} + \frac{\partial \tilde{\psi}^\star}{\partial Z^\star}\right)\right], \end{aligned} \tag{4.29a}$$

$$\gamma^\star \frac{1}{g^\star}\frac{\partial g^\star}{\partial t^\star} = 2\left(1 + \frac{1}{\beta_1}(p^\star - 1)\right)\left[\bar{\mu}_p^\star\left(\frac{p^\star}{g^\star}\right)^2 - \bar{\mu}_G^\star (g^\star)^2\right], \tag{4.29b}$$

where

$$\begin{aligned} T_{zz}^{sf\star} &= \left(1 + \frac{1}{\beta_1}(p^\star - 1)\right)\left[2\bar{\mu}_p^\star\left(\frac{p^\star}{g^\star}\right)^2 + \bar{\mu}_p^\star\left(\frac{(p^\star)^2}{(g^\star)^2} - 1\right) + \bar{\mu}_G^\star\left((g^\star)^2 - 1\right)\right] \\ &\quad + \left(1 + \frac{1}{p^\star}(\beta_1 - 1)\right)p^\star\left[\ln\left(1 - \frac{1}{p^\star}\right) + \frac{1}{p^\star} + \chi\frac{1}{(p^\star)^2}\right], \end{aligned} \tag{4.30a}$$

$$\tilde{\psi}^\star = \frac{\bar{\mu}_G^\star p^\star}{\beta_1}\left((g^\star)^2 - 1\right) + \frac{p^\star \bar{\mu}_p^\star}{\beta_1}\left(\frac{(p^\star)^2}{(g^\star)^2} - 1\right) + p^\star\left[\left(1 - \frac{1}{p^\star}\right)\ln\left(1 - \frac{1}{p^\star}\right) - \chi\frac{1}{(p^\star)^2}\right]. \tag{4.30b}$$

The boundary conditions Eq. (4.27) reduce to

$$-F^\star(t^\star) = T_{zz}^{sf\star}, \quad Z^\star = \pm H^\star, \tag{4.31}$$

with $F^\star = \frac{F}{\mu}$, $P_\infty^\star = \frac{P_\infty^\star}{\mu}$. If we pick $H$ as the characteristic length $L$, then $H^\star = 1$. The coupled equations Eqs. (4.29a), (4.29b) are solved using the staggered scheme shown in algorithm 1.

The ratio of the mass of the swollen solid to its original unswollen mass can be calculated as follows:

$$\begin{aligned}\frac{m}{m_0} &= \frac{\int \rho \, dV}{\int \rho_R^s \, dV} \\ &= \frac{\int_{z=-1}^{z=1} \frac{\rho}{\rho_R^s} \, dz}{\int_{z=-1}^{z=1} dz} \\ &= \frac{1}{2\beta_1}\int_{z=-1}^{z=1}\left[1 + \frac{1}{p^\star}(\beta_1 - 1)\right] dz. \end{aligned} \tag{4.32}$$

Once, the value of $p^\star(Z^\star, t^\star)$ is evaluated on the domain at various times, Eq. (4.32) is integrated numerically to get the mass ratio.

**Algorithm 1** A staggered procedure for solving the coupled equations

Input: $\beta_1$, $\beta_2$, $\chi$, $F^\star$, $P^\star_\infty$, $\bar{\mu}^\star_p$, $\bar{\mu}^\star_G$, $\gamma^\star$; Time of integration $t^\star_f$; Time step $\Delta t^\star$; No. of divisions along $Z^\star$ direction $N$; TOLERANCE.
Output: $\boldsymbol{p}^\star$.
Set $\boldsymbol{p}^{\star 0} = 1$, $\boldsymbol{g}^{\star 0} = 1$.
**while** $t < t_f$ **do**
  $t^\star = t^\star + \Delta t^\star$.
  **while** true **do**
    Using $\boldsymbol{p}^{\star i(l)}$, $\boldsymbol{g}^{\star i(l)}$, $N$, $\Delta t$, $\beta_1$, $\beta_2$, $\chi$, $F^\star$, $P^\star_\infty$, $\bar{\mu}^\star_p$, $\bar{\mu}^\star_G$, $\gamma^\star$, Solve for $\boldsymbol{p}^{\star i(l+1)}$ using Eq. (4.29a) in MATLAB's `pdepe` solver with the $\boldsymbol{p}^{\star i(l+1)}$ at the boundaries ($Z^\star = \pm 1$) obtained by solving the non-linear algebraic equation in Eq. (4.31). This non-linear algebraic equation is solved using the `fsolve` solver in MATLAB.
    Using $\boldsymbol{p}^{\star i(l+1)}$, $\boldsymbol{g}^{\star i(l)}$ and $\Delta t$, $\beta_1$, $\beta_2$, $\chi$, $F^\star$, $P^\star_\infty$, $\bar{\mu}^\star_p$, $\bar{\mu}^\star_G$, $\gamma^\star$, Solve for $\boldsymbol{g}^{\star i(l)}$ using Eq. (4.29b) in MATLAB's `ode45` solver.
    **if** $\|\boldsymbol{p}^{\star i(l+1)} - \boldsymbol{p}^{\star i(l)}\|_2 <$ TOLERANCE **then**
      Return.
    **end if**
  **end while**
  $\boldsymbol{p}^{\star i+1} \leftarrow \boldsymbol{p}^{\star i}$, $\boldsymbol{g}^{\star i+1} \leftarrow \boldsymbol{g}^{\star i}$.
**end while**

4.3. **Comparison with experimental data.** Fig. (3) shows comparison of the numerical results to the experimental data for the ratio of swollen to unswollen (see Eq. (4.32)) PMDA-ODA (poly(N, N'- bisphenoxyphenylpyromellitimide)) due to diffusion of the solvents DMSO (dimethylsulfoxide) and NMP (N-methyl-2-pyrollidinone). In case of DMSO diffusing through PMDA-ODA the following material parameters were chosen: Density of DMSO was chosen to be $1.096\, g/cc$ (Yang et al. (2008)) and density of PMDA-ODA to be $1.42\, g/cc$ (Srinivasan et al. (1994)), and so $\beta_1 = 1.3$. Also, $\chi = 0.425$, $\beta_2 = 0.018$, $\bar{\mu}^\star_p = 0.1$, $\bar{\mu}^\star_G = 0.1$, $\gamma^\star = 20$ were chosen. The characteristic time ($T$) chosen was $10500\, min$. For the diffusion of NMP, the material parameters chosen were: density of NMP is taken to be $1.02\, g/cc$ (Yang et al. (2008)) and so $\beta_1 = 1.4$. Next, $\chi = 0.6$, $\beta_2 = 0.016$, $\bar{\mu}^\star_p = 0.1$, $\bar{\mu}^\star_G = 0.1$, $\gamma^\star = 20$ and the characteristic time chosen was $245\, min$. The numerical results show good agreement with the experimental data taken from (Gattiglia and Russell (1989)).

Next, we shall consider the diffusion of water through HPFE-II-52. The material parameters were assumed to be: $\chi = 0.425$, $\beta_1 = 1.3$, $\beta_2 = 0.018$, $\bar{\mu}^\star_p = 0.1$, $\bar{\mu}^\star_G = 0.1$, $\gamma^\star = 20$. The characteristic time chosen was $2800\, s$. Even in this case the numerical results and experimental data taken from (Bhargava (2007)) match well (see Fig. (4)). In all the numerical calculations TOLERANCE was chosen to be $10^{-4}$.

We shall now consider the problem of compression of the viscoelastic solid and study its effects on swelling due to diffusion of a fluid. In this numerical experiment, the solid is allowed to swell freely first till it saturates with fluid (upto $t^\star = 0.5$). Then, the swollen solid is subjected to constant compressive force of $F^\star = 1$ is applied for a time period of $t^\star = 0.5$ and then the load is removed, and the solid is allowed to swell freely again for another time period of $t^\star = 0.5$. Fig. (5a) shows that the volume of the solid gradually increases with time and then reaches a steady state where the volume of the solid is same everywhere and there

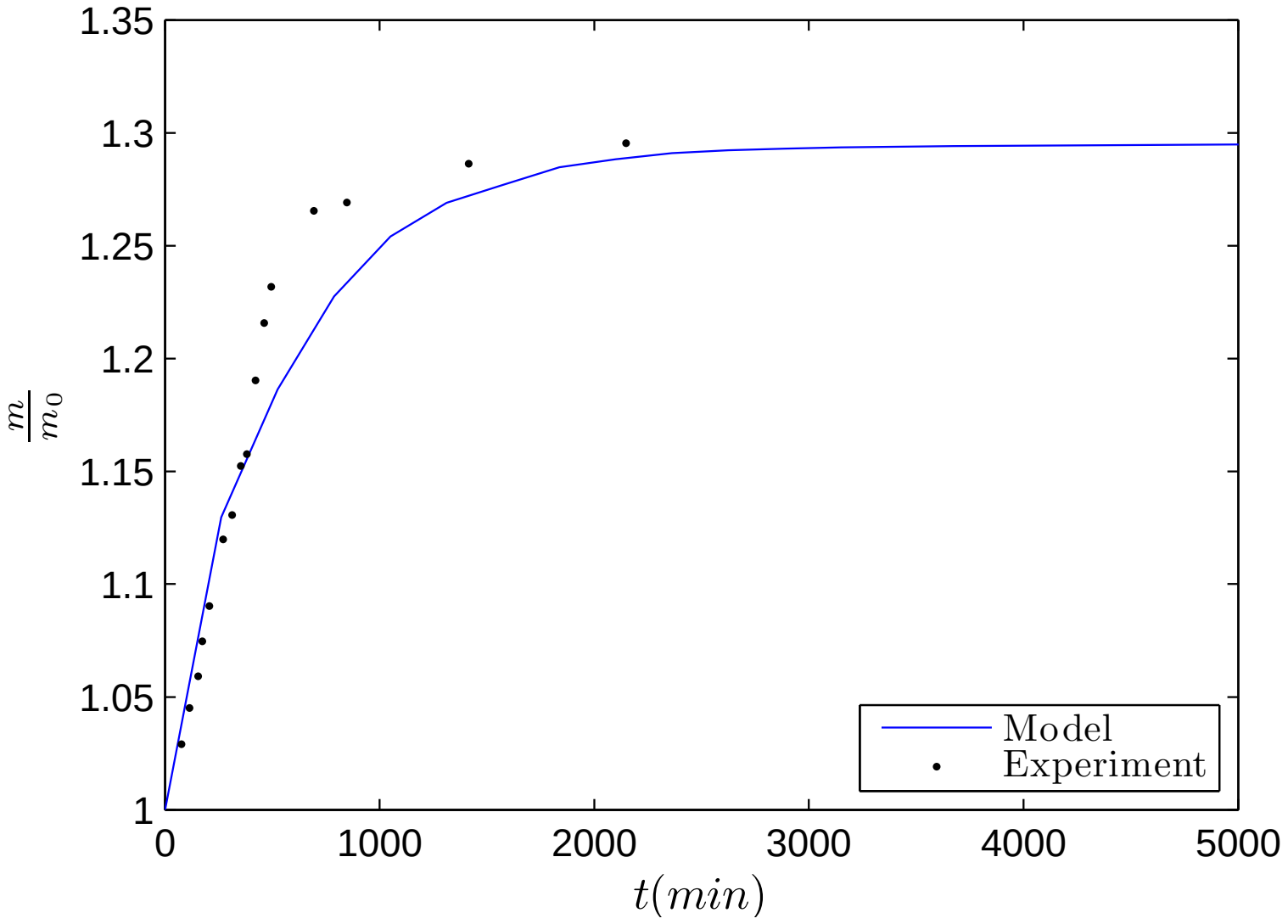

(A) DMSO

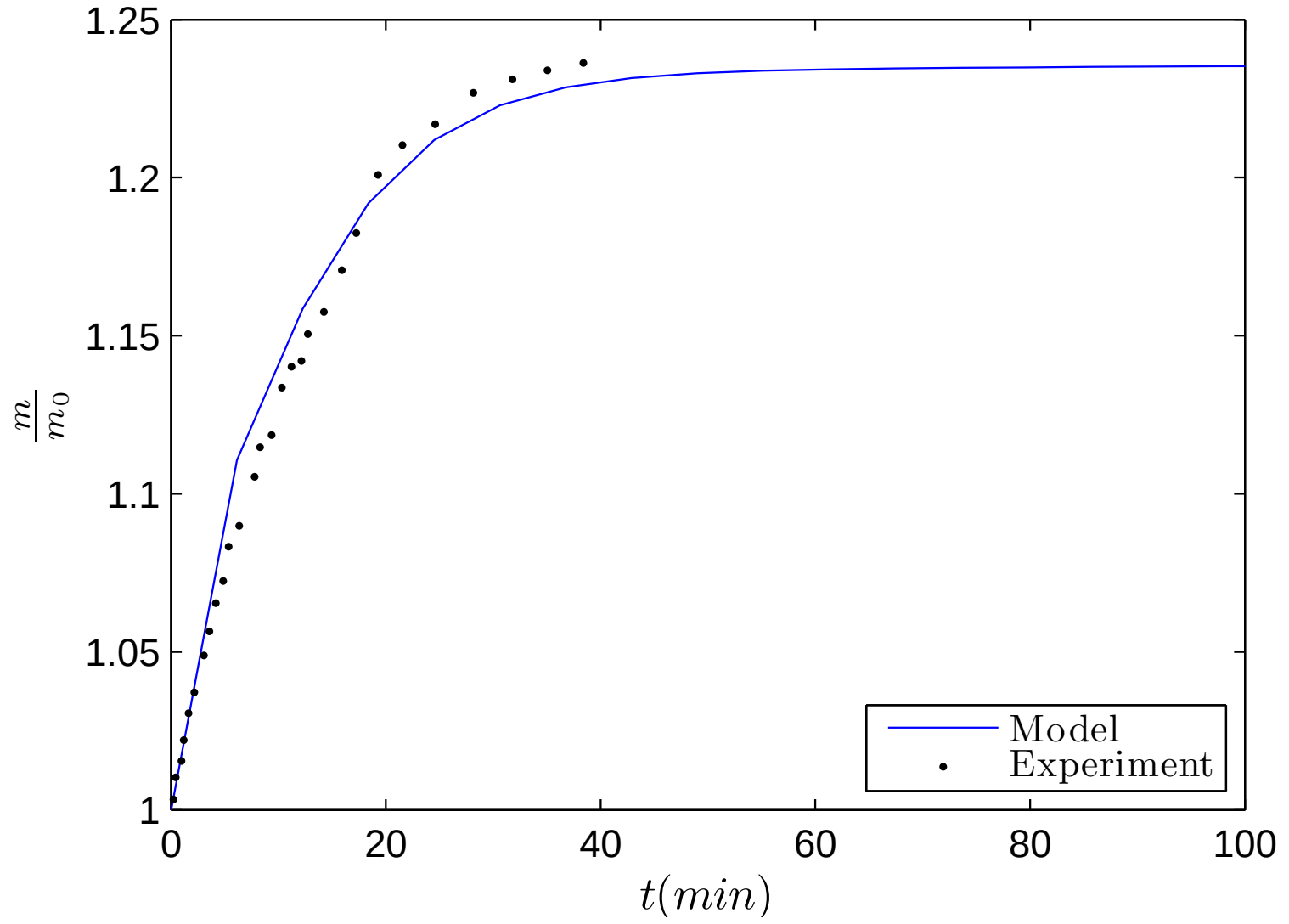

(B) NMP

FIGURE 3. Comparison of the model with the experimental data from (Gattiglia and Russell (1989)) for the diffusion of DMSO and NMP through PMDA-ODA (imidized at $300^o$C) under free-swelling condition. The characteristic times chosen were 10500 $min$ and 245 $min$ for DMSO and NMP, respectively. Here, 301 spatial points were used for the calculations, non-dimensional time step chosen is $\Delta t^\star = 0.025$.

is no further swelling. Also, the volume of the solid near the boundary increases faster than that of the inner solid. Upon application of a constant compressive load in Fig. (5b), the fluid diffuses out of the swollen solid and the volume of the solid gradually decreases until the volume of the solid is same everywhere. Next, upon removal of the compressive load in

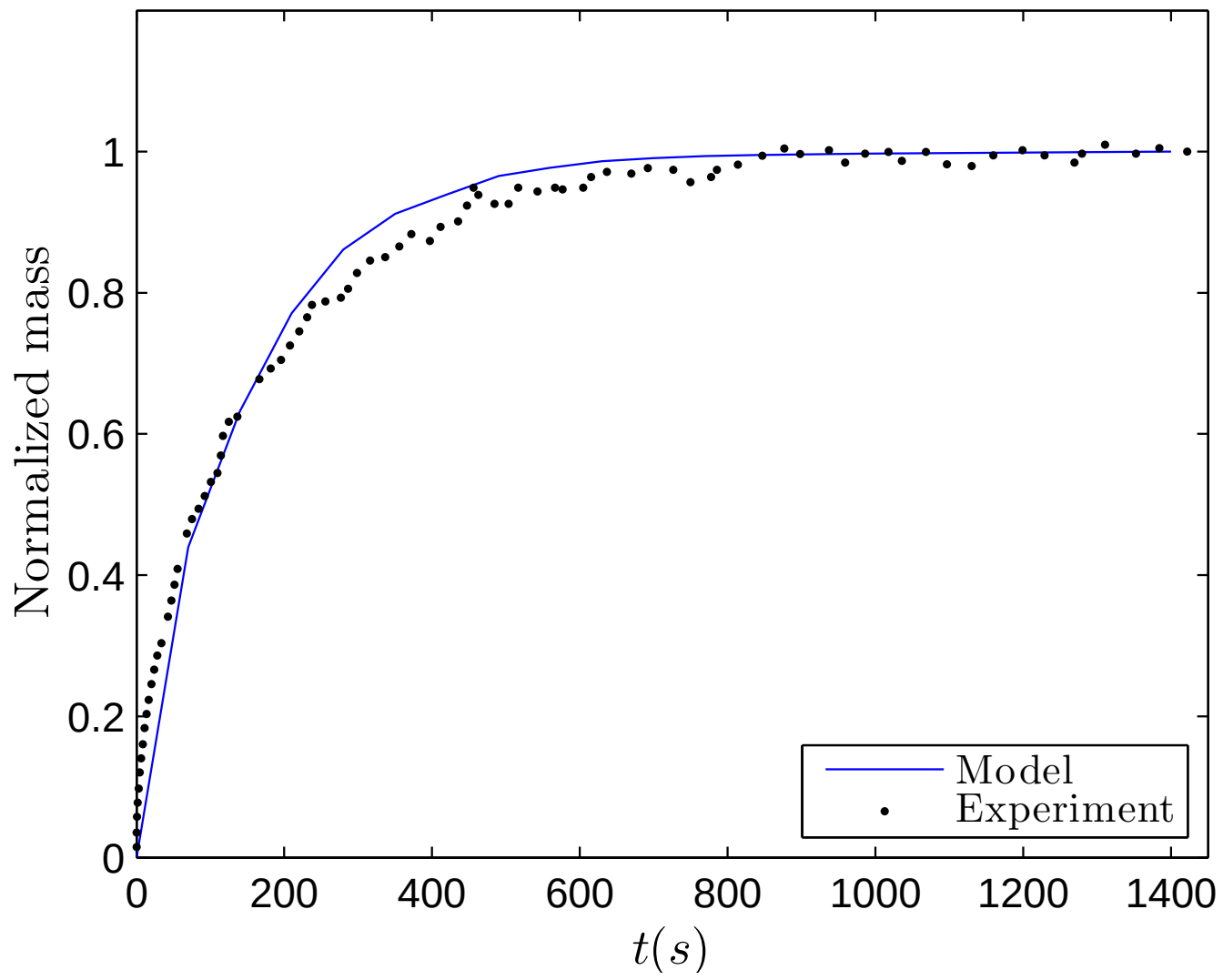

FIGURE 4. Comparison of the model with the experimental data from (Bhargava (2007)) (pg. 27) for the diffusion of water through HFPE-II-52 under free-swelling condition. The characteristic time chosen was $2800\,s$. The parameters chosen are $\chi = 0.425$, $\beta_1 = 1.3$, $\beta_2 = 0.018$, $\bar{\mu}_p^\star = 0.1$, $\bar{\mu}_G^\star = 0.1$, $\gamma^\star = 20$. Here, 301 spatial points were used for the calculations, $\Delta t^\star = 0.025$. The normalized mass is defined by $\dfrac{m(t) - m_0}{m_\infty - m_0}$, where $m_0$ is the mass of the dry solid, $m_\infty$ is the steady state mass of the swollen solid, $m(t)$ is the mass of the swollen solid at a given time $t$.

Fig. (6), the solid absorbs the fluid and swells freely back to its original swollen saturation state.

## 5. Conclusions

We developed a systematic framework with a thermodynamic basis to develop constitutive relations for the diffusion of a fluid through a viscoelastic solid. A model was also derived using this framework by choosing specific forms for the Helmholtz potential and the rate of dissipation, and by maximizing the rate of dissipation. An initial boundary value problem was solved where we considered free swelling and swelling under the application of external force. We also showed that the model fits well with the experimental data for free swelling of polyimides. Furthermore, our work in this paper can be easily extended to study the diffusion of fluids in biological materials as well as in studying moisture-induced damage in asphalt mixes and other geomaterials that show viscoelastic behavior.

Finally, here we shall summarize the assumptions made in this paper:

(1) the specific Helmholtz potential of the constituents is the same,
(2) the temperature of the constituents is the same,
(3) the specific Helmholtz potential of the mixture depends on the temperature of the mixture, and the deformation of the solid,

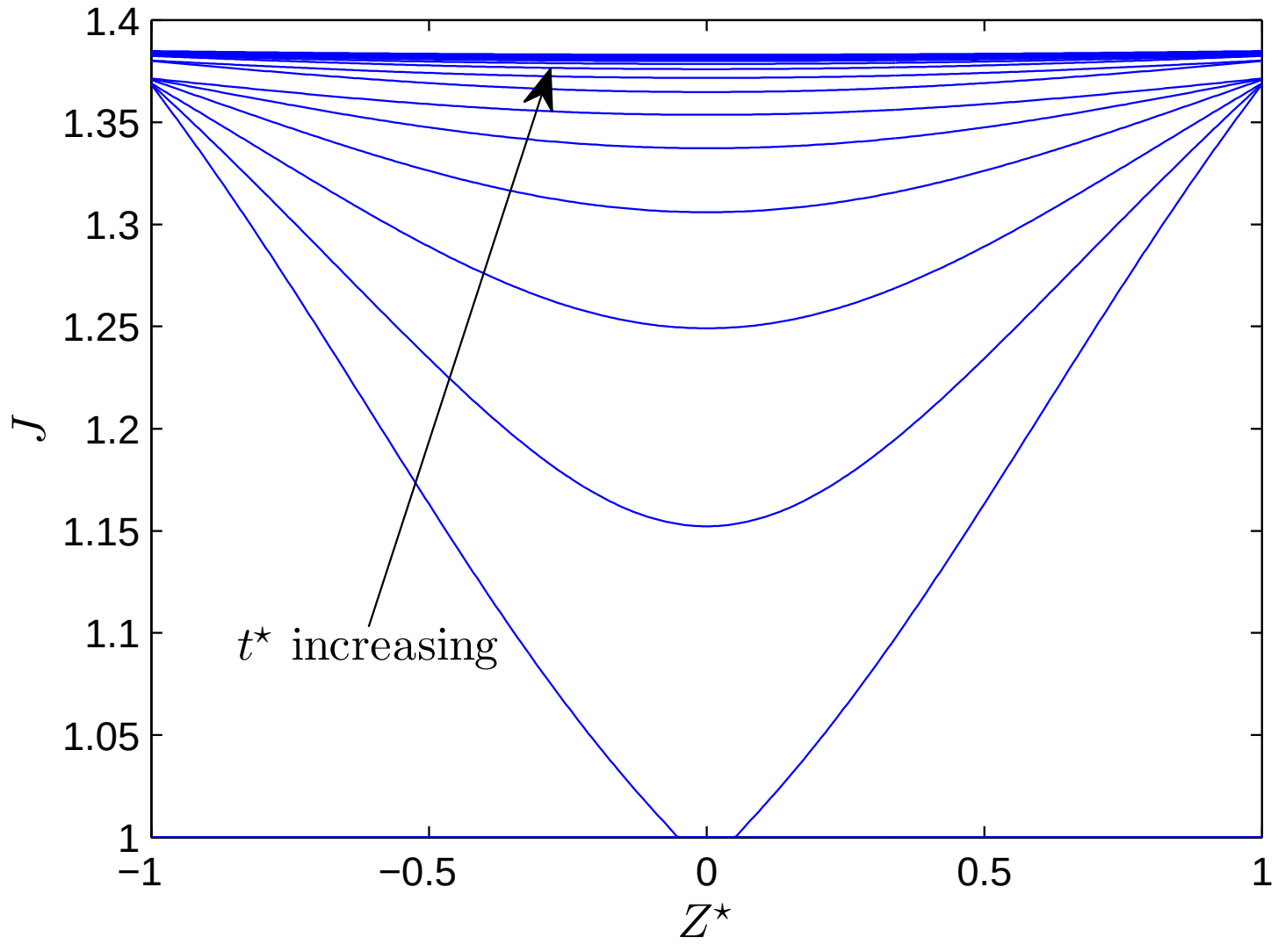

(A) free swelling

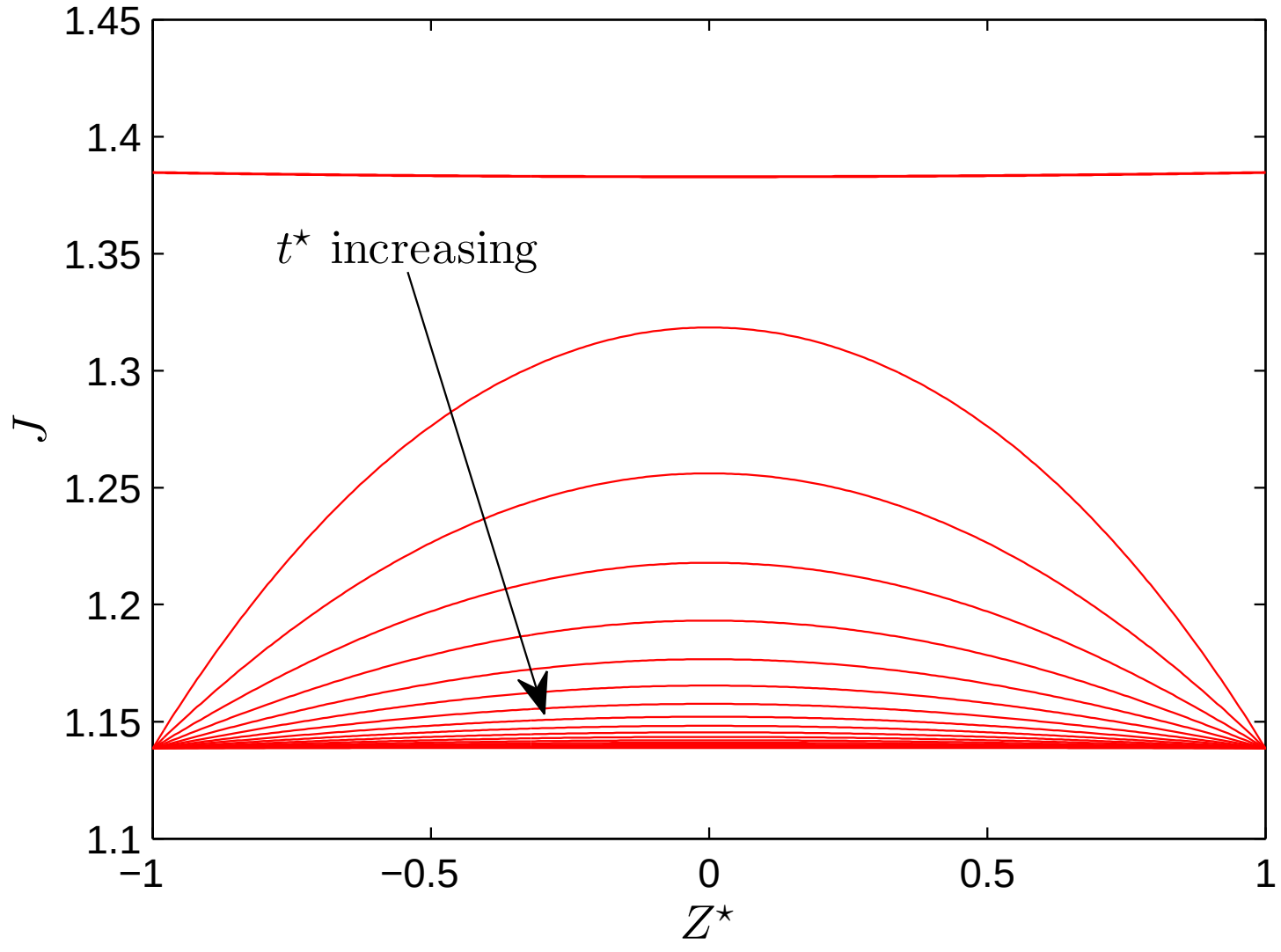

(B) compression after free swelling

FIGURE 5. Ratio of volume of the swollen solid to volume of unswollen solid ($J = \det(\boldsymbol{F}^s)$) as a function of time for (a) free swelling and (b) compressive force $F^\star = 1$ is applied after the swollen solid reaches a saturated state due to free swelling. The parameters chosen are $\chi = 0.425$, $\beta_1 = 1.3$, $\beta_2 = 0.018$, $\bar{\mu}_p^\star = 0.1$, $\bar{\mu}_G^\star = 0.1$, $\gamma^\star = 20$. Here, 301 spatial points were used for the calculations, non-dimensional time step chosen is $\Delta t^\star = 0.025$.

(4) the volume of the mixture is sum of the volumes of the constituents in their natural state,
(5) the response of the solid from the current configuration to its natural configuration is isotropic and elastic,

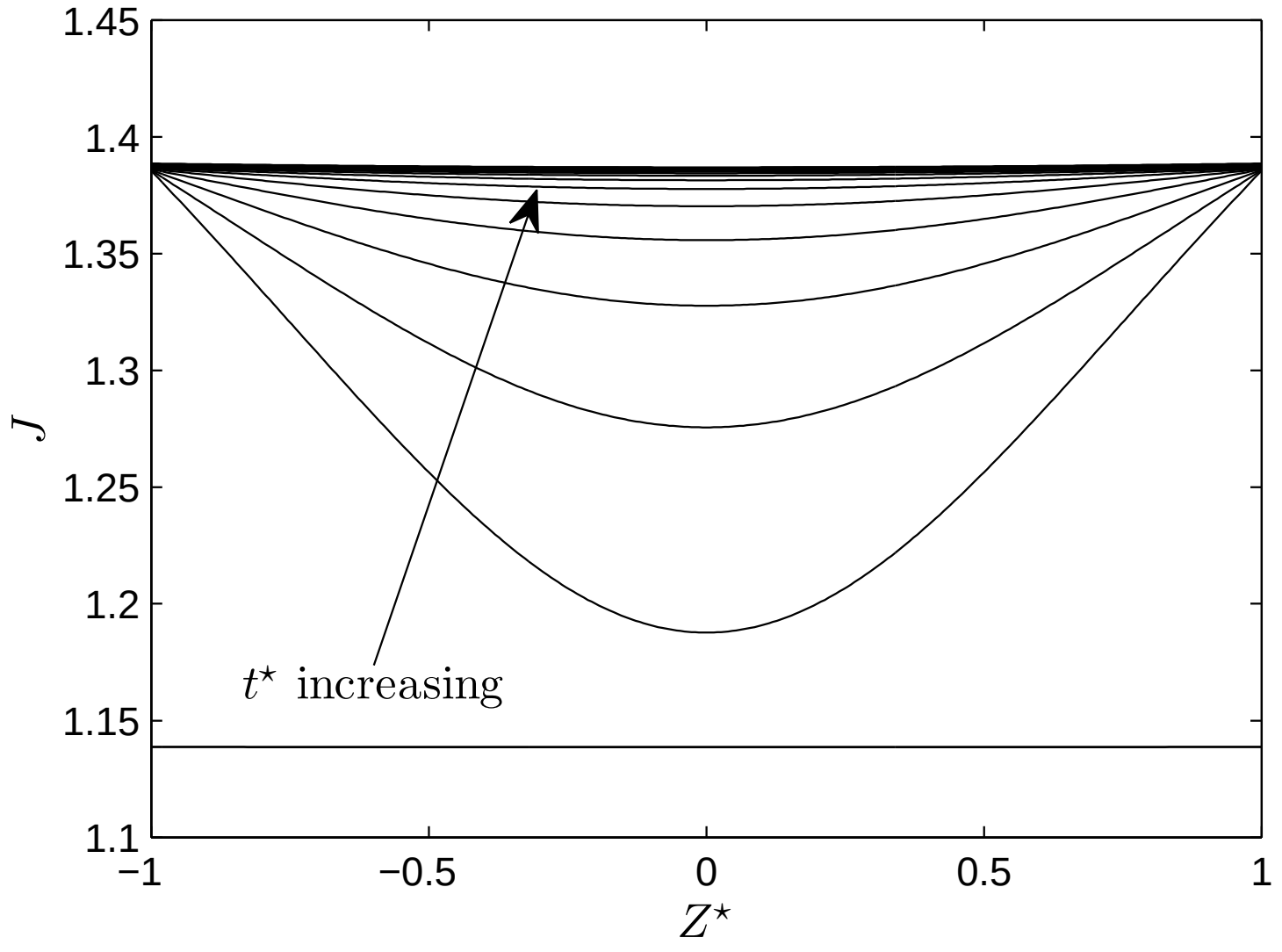

Figure 6. Ratio of volume of the swollen solid to volume of unswollen solid ($J = \det(\boldsymbol{F}^s)$) as a function of time for free- swelling after the compressive force is removed. The parameters chosen are $\chi = 0.425$, $\beta_1 = 1.3$, $\beta_2 = 0.018$, $\mu_p^\star = 0.1$, $\mu_G^\star = 0.1$, $\gamma^\star = 20$. Here, 301 spatial points were used for the calculations, non-dimensional time step chosen is $\Delta t^\star = 0.025$.

(6) the reference configurations of the constituents are same as their natural states,
(7) the rate of dissipation of the mixture is assumed to be the sum of the rates of dissipation due to mechanical working of the viscoelastic solid, due to the fluid (i.e, due to the friction between the layers of the fluid), and due to the drag between the solid and the fluid.

The following additional assumptions are made to solve the problem of compression:

(1) the viscosity of the fluid is zero i.e., we are assuming that the dissipation due to the friction between the layers of the fluid is much smaller than that due to the drag between the solid and the fluid,
(2) the velocity and acceleration of the solid are small compared to that of the fluid,
(3) acceleration of the fluid is small,
(4) the external loading is applied on the solid only, whereas the fluid pressure at the boundary is borne by both the solid and the fluid.

## Acknowledgements

The author would like to thank AFOSR/AFRL for their support. Part of this work has been done when the author was appointed as a lecturer during his Ph.D. by the Department of Mechanical Engineering at Texas A&M University. This support by the department is appreciated. The author would also like to thank Mr. Shriram Srinivasan and Dr. K. B. Nakshatrala of TAMU for their help.

## Appendix A. Convergence of numerical results

Since the analytical solution for the problem is unknown, we perform an *engineering* convergence study of the solution using the described algorithm (see figure (7)). In this study, the error is calculated by taking the difference between solution of various grid sizes (5, 15, 25, . . . , upto 351 points) and the solution found using a very fine grid of 401 points at the point $Z^\star = 0$ and at a time of $t^\star = 0.5$. Note that the error is propotional to logarithm of the spatial increment and hence the convergence rate is slow. The aim of our current work is not to present an optimal algorithm but to solve the coupled partial differential equations.

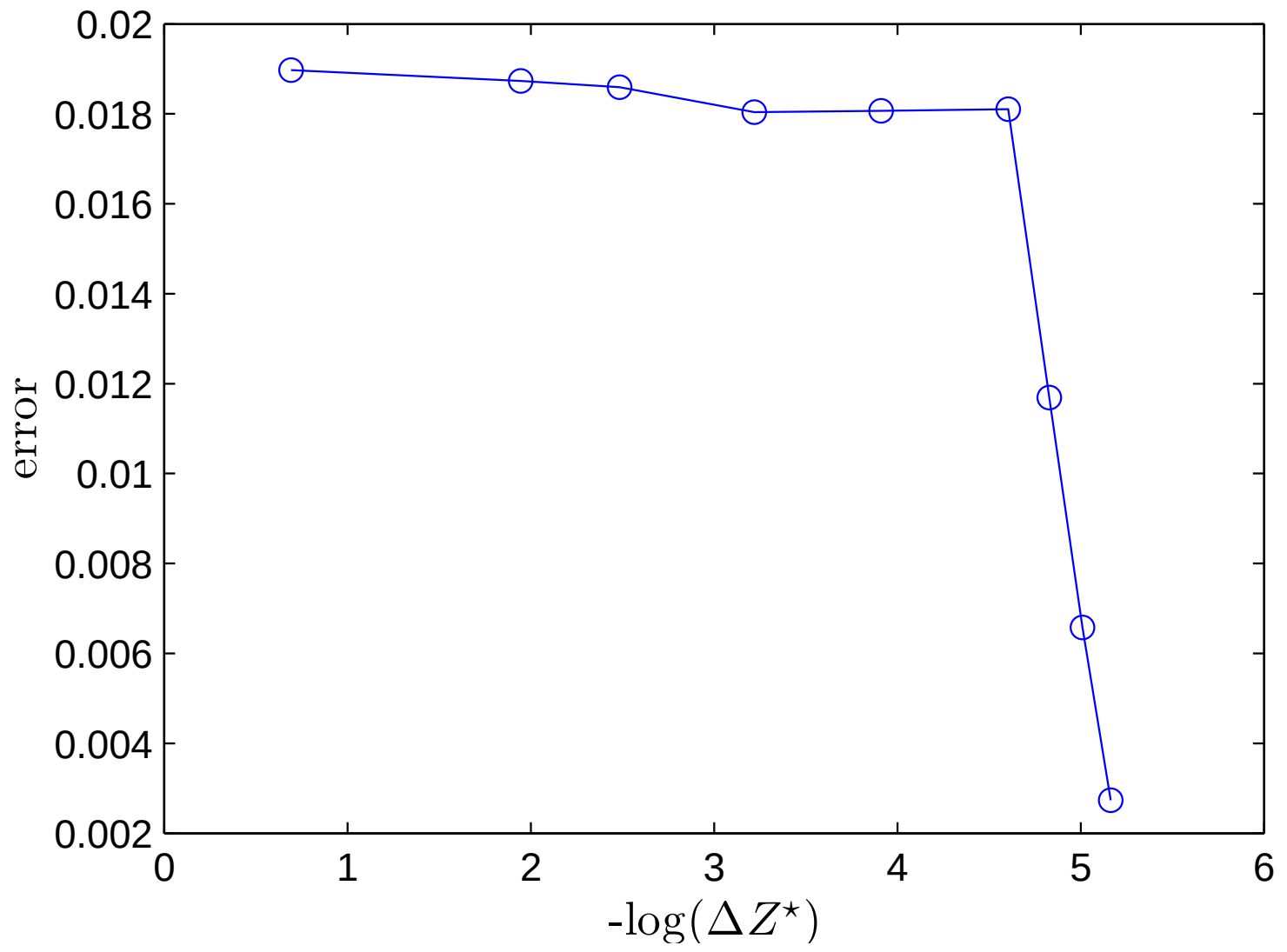

FIGURE 7. Engineering spatial convergence of the solution for $p$ at $Z^\star = 0$ and at $t^\star = 0.5$. The time-step was chosen to be $\Delta t^\star = 0.025$.

## APPENDIX B. DERIVATION OF THE CONSTITUTIVE EQUATION FOR THE VISCOELASTIC SOLID IN THE ABSENCE OF DIFFUSION

Here we show that in the absence of diffusion the derived constitutive equations reduces to a variant of the three-dimensional standard linear solid model given in (Karra and Rajagopal (2010)). Now, in the absence of diffusion of the fluid, we will have to drop the last term in (3.33) to get

$$\hat{\psi} = A^s + \left(B^s + c_2^s\right)\left(\theta - \theta_s\right) - \frac{c_1^s}{2}\left(\theta - \theta_s\right)^2 - c_2^s\theta \ln\left(\frac{\theta}{\theta_s}\right) + \frac{\mu_{G0} - \mu_{G1}\theta}{\rho^s\theta_s}(\mathrm{I}_{B_G^s} - 3) + \frac{\mu_{p0} - \mu_{p1}\theta}{\rho^s\theta_s}(\mathrm{I}_{B^s_{p(t)}} - 3). \tag{B.1}$$

We shall assume that the solid is incompressible in the absence of fluid. The constraint of incompressibility is given by

$$\mathrm{tr}\left(\boldsymbol{D}^s\right) = \mathrm{tr}\left(\boldsymbol{D}_G^s\right) = 0. \tag{B.2}$$

The reduced energy dissipation equation of the solid reduces to

$$\boldsymbol{T}^s \cdot \boldsymbol{D}^s - \left(\rho\frac{d\psi}{dt}\right)_{\theta\ \mathrm{fixed}} = \xi_m. \tag{B.3}$$

In the absence of diffusion, there will be only be dissipation due to mechanical working of the solid, and so the rate of dissipation in this case would be

$$\xi_m = \gamma(\theta)\boldsymbol{D}_G^s \cdot \boldsymbol{D}_G^s. \tag{B.4}$$

Upon maximizing the rate of dissipation using (B.4), (B.2) as constraints (see (Karra and Rajagopal (2010)) for further details) we arrive at

$$\boldsymbol{T}^s = p\boldsymbol{I} + 2\bar{\mu}_p \boldsymbol{B}^s_{p(t)}, \tag{B.5a}$$

$$\boldsymbol{T}^s = \lambda\boldsymbol{I} + 2\bar{\mu}_G \boldsymbol{B}^s_G + \eta\boldsymbol{D}^s_G. \tag{B.5b}$$

From (B.5a), (B.5b) we have

$$(p-\lambda)\,\boldsymbol{I} + 2\bar{\mu}_p \boldsymbol{B}^s_{p(t)} - 2\bar{\mu}_G \boldsymbol{B}^s_G = \eta\boldsymbol{D}^s_G. \tag{B.6}$$

Taking the trace of (B.6), we get

$$3\,(p-\lambda) = -2\bar{\mu}_p \mathrm{tr}\left(\boldsymbol{B}^s_{p(t)}\right) + 2\bar{\mu}_G \mathrm{tr}\left(\boldsymbol{B}^s_G\right). \tag{B.7}$$

Hence, (B.7) in (B.6) gives

$$2\bar{\mu}_p \boldsymbol{B}^s_{p(t)} - 2\bar{\mu}_G \boldsymbol{B}^s_G = \frac{2}{3}\left[\bar{\mu}_p \mathrm{tr}\left(\boldsymbol{B}^s_{p(t)}\right) - \bar{\mu}_G \mathrm{tr}\left(\boldsymbol{B}^s_G\right)\right] + \eta\boldsymbol{D}^s. \tag{B.8}$$

The final constitutive equations for the viscoelastic solid are

$$\boldsymbol{T}^s = p\boldsymbol{I} + 2\bar{\mu}_p \boldsymbol{B}^s_{p(t)}, \tag{B.9}$$

with (B.8) being the evolution equation for the natural configuration of the viscoelastic solid. This is a variant of the model derived in (Karra and Rajagopal (2010)).

Satish Karra, Texas A&M University, Department of Mechanical Engineering, 3123 TAMU, College Station TX 77843-3123, United States of America

*E-mail address*: `satkarra@tamu.edu`